\pdfoutput=1

\documentclass[11pt,twoside,a4paper,cmspaper,final,collab]{cms-tdr}
\def\svnVersion{493190P}\def\cmsCernNoTag{CERN-EP-2018-311}\def\cmsCernDate{\today}\def\cmsMessage{Published in the Journal of High Energy Physics as \href{http://dx.doi.org/10.1007/JHEP03(2019)141}{\doi{10.1007/JHEP03(2019)141.}}}

\begin{document}\cmsNoteHeader{EXO-18-010}

\hyphenation{had-ron-i-za-tion}
\hyphenation{cal-or-i-me-ter}
\hyphenation{de-vices}
\RCS$HeadURL: svn+ssh://svn.cern.ch/reps/tdr2/papers/EXO-18-010/trunk/EXO-18-010.tex $
\RCS$Id: EXO-18-010.tex 490797 2019-03-03 17:32:59Z alverson $

\newlength\cmsTabSkip\setlength{\cmsTabSkip}{2ex}
\providecommand{\cmsTable}[1]{\resizebox{\textwidth}{!}{#1}}

\cmsNoteHeader{EXO-18-010}
\title{Search for dark matter produced in association with a single top quark or a top quark pair in proton-proton collisions at $\sqrt{s}=13\TeV$}

\newcommand{\ttjets}{\ensuremath{\ttbar}+jets\xspace}
\newcommand{\tttwol}{\ensuremath{\ttbar(2\ell)}\xspace}
\newcommand{\ttonel}{\ensuremath{\ttbar(1\ell)}\xspace}
\newcommand{\wjets}{\ensuremath{\PW}+jets\xspace}
\newcommand{\zjets}{\ensuremath{\PZ}+jets\xspace}
\newcommand{\Zinv}{\ensuremath{{\cPZ} \to \nu \nu}\xspace}
\newcommand{\ZLL}{\ensuremath{{\cPZ} \to \ell \ell}\xspace}
\newcommand{\mchi}{\ensuremath{{m}_{\chi}}\xspace}
\newcommand{\tDM}{\ensuremath{{\cPqt}/\overline{{\cPqt}}}+DM\xspace}
\newcommand{\ttDM}{\ensuremath{\ttbar}+DM\xspace}
\newcommand{\bbDM}{\ensuremath{\bbbar}+DM\xspace}
\newcommand{\tttDM}{\ensuremath{{\cPqt}}, \ensuremath{\ttbar}+DM\xspace}
\newcommand{\njet}{\ensuremath{n_{\text{jet}}}\xspace}
\newcommand{\nlep}{\ensuremath{n_{\text{lep}}}\xspace}
\newcommand{\nbjet}{\ensuremath{n_{\PQb}}\xspace}
\newcommand{\HTratio}{\ensuremath{\pt(\mathrm{j_{1})}/\HT}\xspace}
\newcommand{\mindphi}{\ensuremath{\text{min}\Delta\phi(\mathrm{j_{1,2}}, \ptvecmiss)}\xspace}
\newcommand{\mTb}{\ensuremath{\mT^{\cPqb}}\xspace}
\newcommand{\mTt}{\ensuremath{\mTii^{\PW}}\xspace}
\newcommand{\mll}{\ensuremath{m_{\ell\ell}}\xspace}

\date{\today}

\abstract{
A search for dark matter produced in association with top quarks in proton-proton collisions at a center-of-mass energy of 13\TeV is presented. The data set used corresponds to an integrated luminosity of 35.9\fbinv recorded with the CMS detector at the LHC.
Whereas previous searches for neutral scalar or pseudoscalar mediators considered dark matter production in association with a top quark pair only, this analysis also includes production modes with a single top quark.
The results are derived from the combination of multiple selection categories that are defined to target either the single top quark or the top quark pair signature. No significant deviations with respect to the standard model predictions are observed. The results are interpreted in the context of a simplified model in which a scalar or pseudoscalar mediator particle couples to a top quark and subsequently decays into dark matter particles. Scalar and pseudoscalar mediator particles with masses below 290 and 300\GeV, respectively, are excluded at 95\% confidence level, assuming a dark matter particle mass of 1\GeV and mediator couplings to fermions and dark matter particles equal to unity.
}

\hypersetup{
pdfauthor={CMS Collaboration},
pdftitle={Search for dark matter produced in association with a single top quark or a top quark pair},
pdfsubject={CMS},
pdfkeywords={CMS, dark matter, top quark}}

\maketitle

\section{Introduction}

Astrophysical observations provide evidence of the existence of nonluminous matter that can be inferred from gravitational effects on galaxies and other large scale objects in the Universe.
While the nature of this dark matter (DM) is still unknown, a compelling candidate is the so-called weakly interacting massive particle~\cite{Bertone:2004pz}.
This new particle is predicted to have weak
interactions with standard model (SM) particles, allowing for direct- and indirect-detection experiments, as well as for searches at collider experiments.

\begin{figure}[htbp!]
\centering
\includegraphics[width=0.28\textwidth]{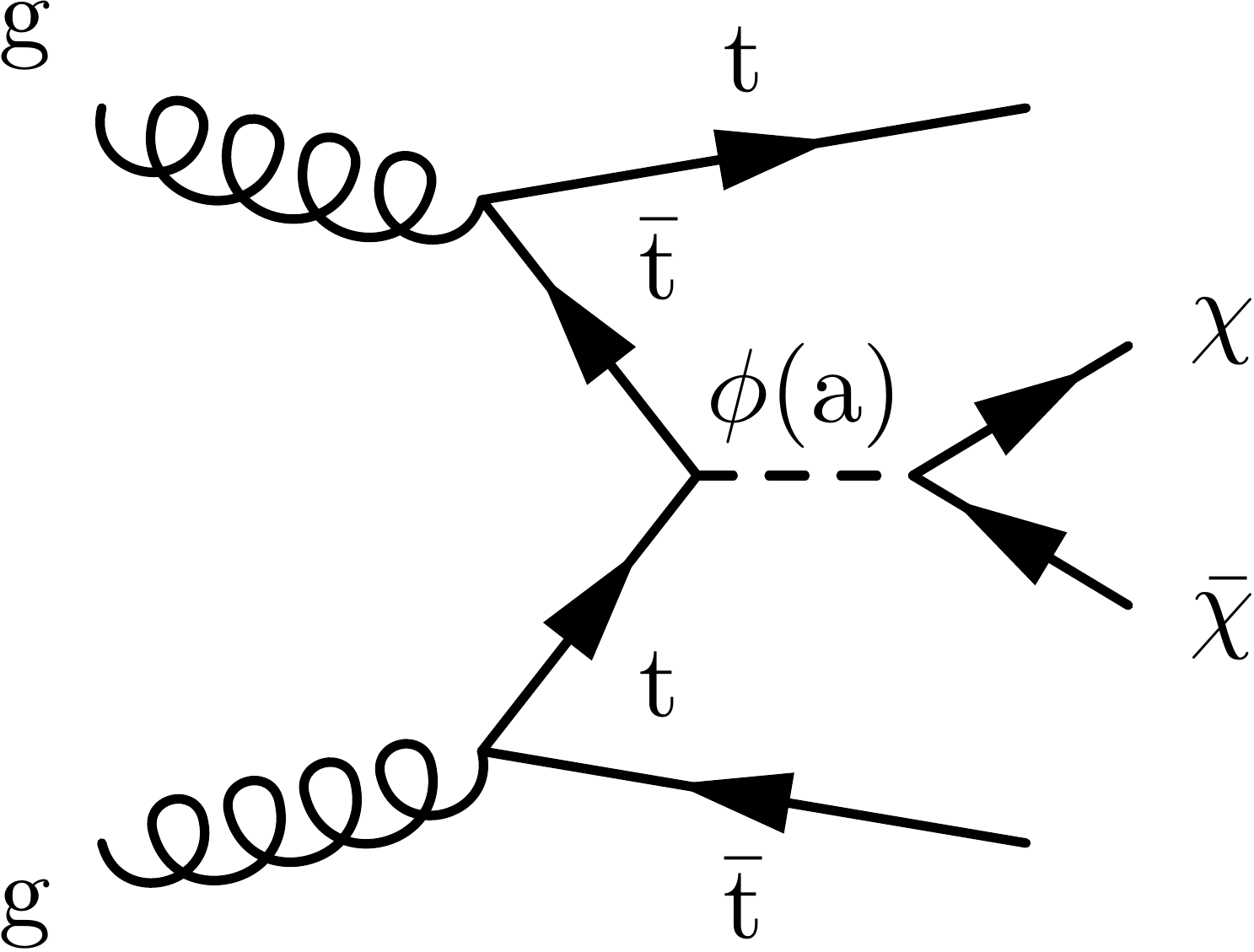}
\hspace{30pt}
\includegraphics[width=0.28\textwidth]{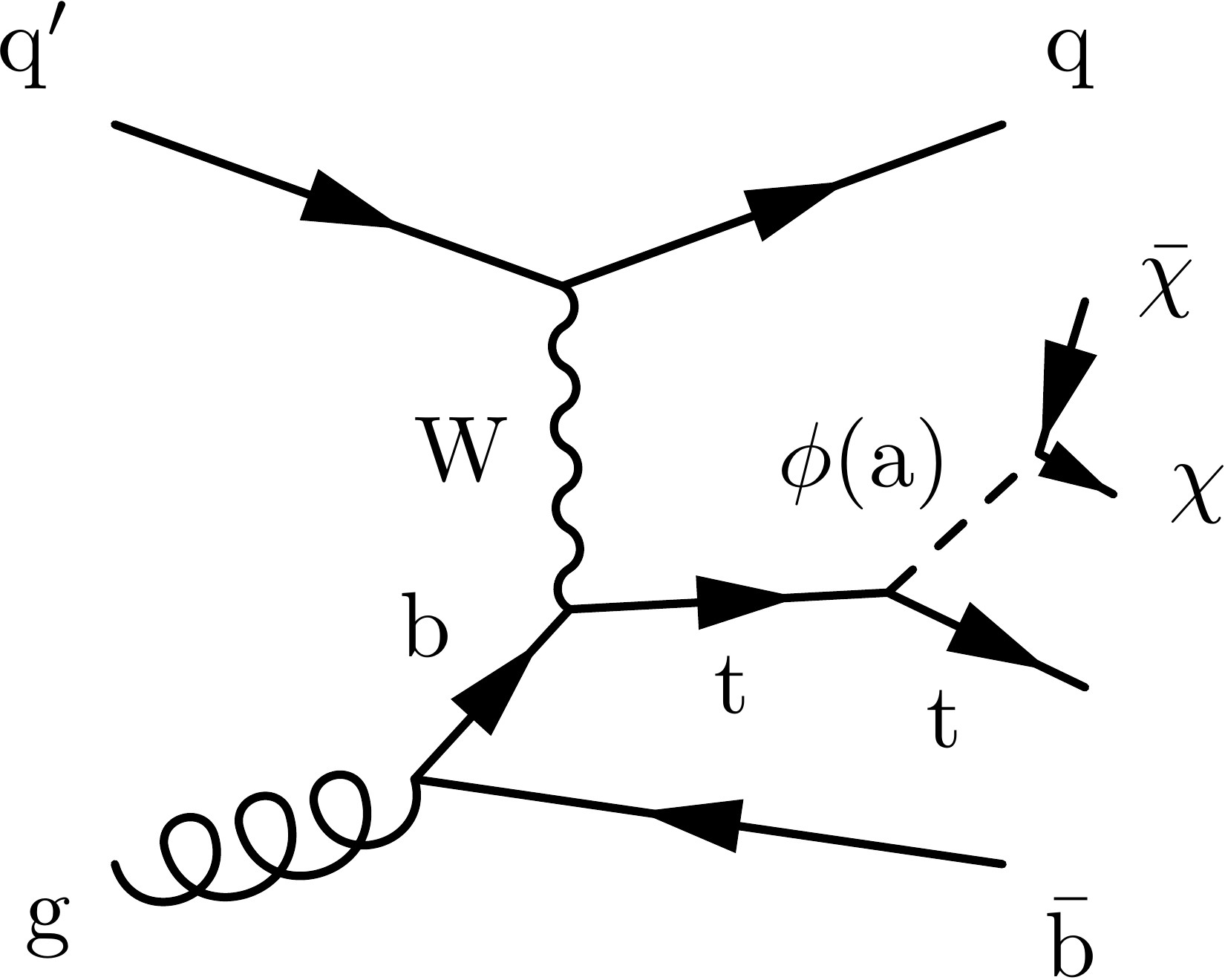}\\[3ex]
\includegraphics[width=0.28\textwidth]{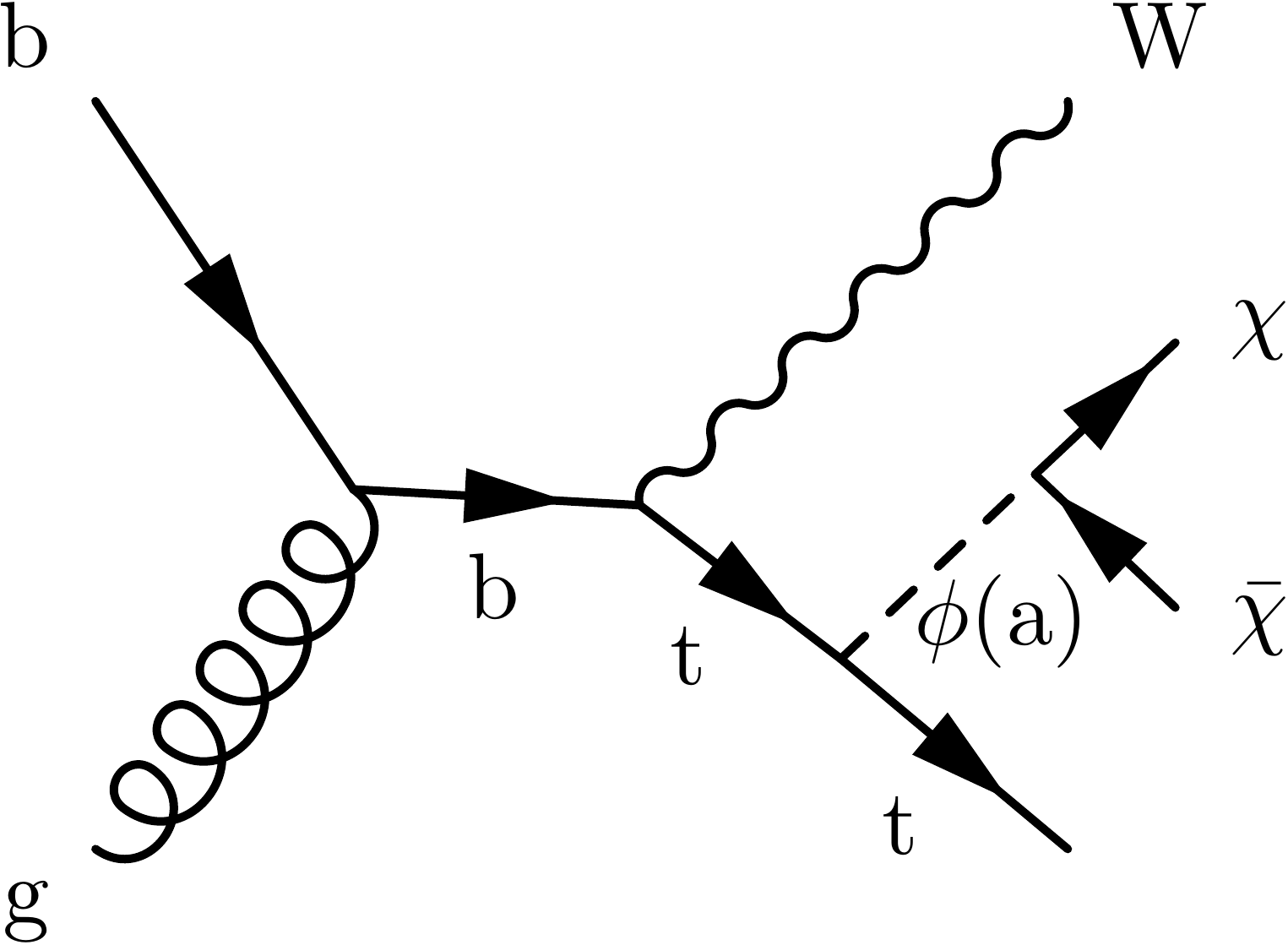}
\hspace{30pt}
\includegraphics[width=0.25\textwidth]{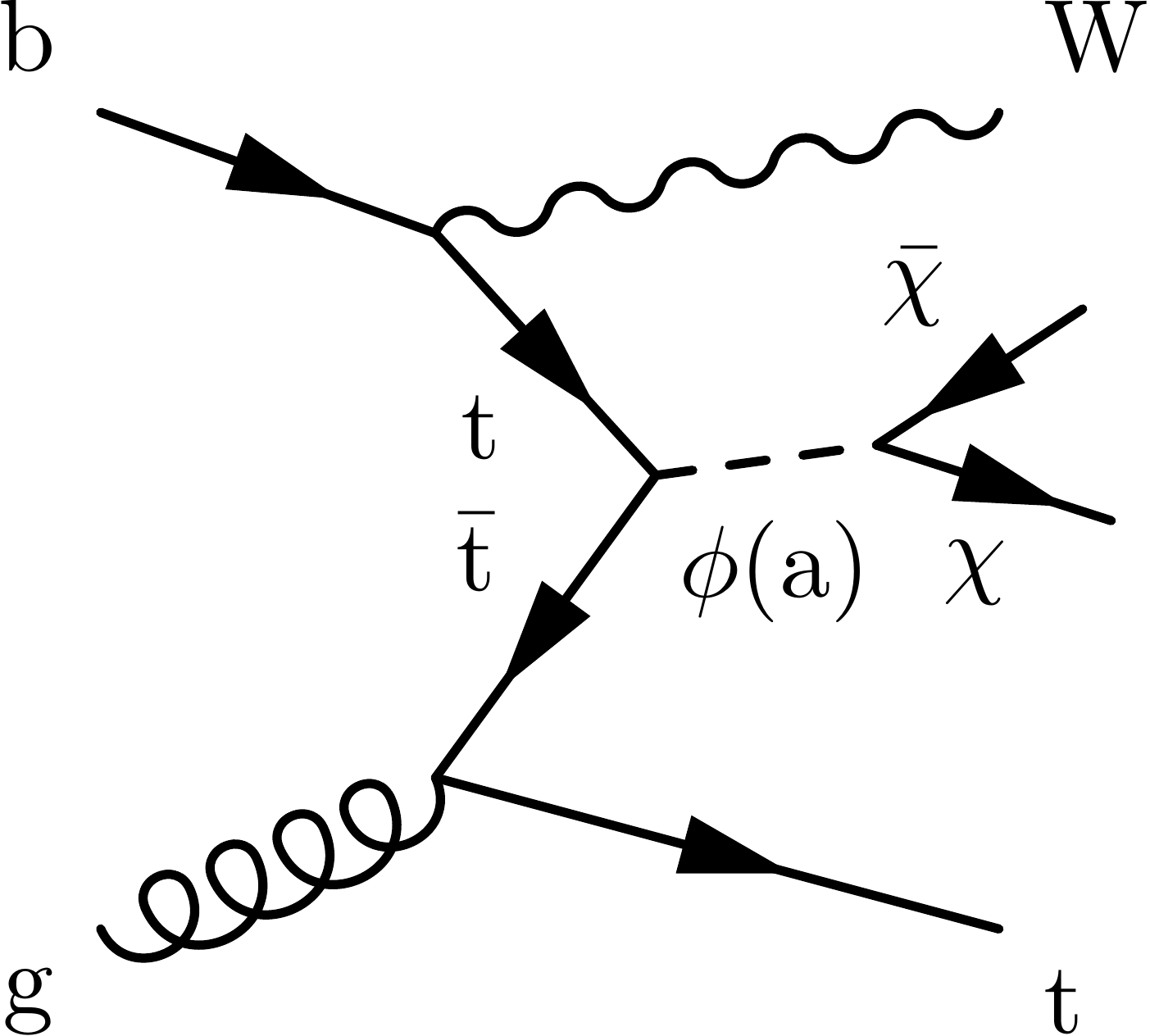}
\caption{Principal production diagrams for the associated production at the LHC of dark matter with a top quark pair (upper left) or a single top quark with associated $t$ channel {\PW} boson production (upper right) or with associated t{\PW} production (lower left and right).}
\label{fig:STdiagram}
\end{figure}

Among all the possible interactions between the SM and DM sectors, it is of particular
interest to investigate interactions mediated by a new neutral scalar or pseudoscalar particle that decays into DM particles, as these can be easily accommodated in models containing extended Higgs boson sectors \cite{Kim:2016csm,extendedHiggs_1,Kim:2008pp,extendedHiggs_2}.
Assuming that this DM scenario respects the principle of minimal flavor violation~\cite{DAmbrosio:2002ex,Isidori:2012ts}, the interactions of this new spin-0 mediator particle follow the same Yukawa coupling structure as in the SM. Therefore, the mediator would couple preferentially to
heavy third-generation quarks.
Assuming the DM particles to be Dirac fermions, the interaction Lagrangian terms for the production of a scalar ($\phi$) or pseudoscalar (\Pa) mediator particle can be expressed as:
\begin{linenomath}
\begin{align}
\mathcal{L}_{\phi} & \supset g_{\chi}\phi\overline{\chi}\chi+\frac{g_{\cPq}\phi}{\sqrt{\smash[b]2}}\sum_{\mathrm{f}}(y_{\mathrm{f}}\overline{\mathrm{f}}\mathrm{f})\label{eq:L_scalar}, \\
\mathcal{L}_{\Pa} & \supset  ig_{\chi}\Pa\overline{\chi}\gamma^{5}\chi+\frac{ig_{\cPq}\Pa}{\sqrt{\smash[b]2}}\sum_{\mathrm{f}}(y_{\mathrm{f}}\overline{\mathrm{f}}\gamma^{5}\mathrm{f}),
\label{eq:L_pseudoscalar}
\end{align}
\end{linenomath}
where the sum runs over the SM fermions $\mathrm{f}$, $y_{\mathrm{f}}=\sqrt{\smash[b]2}m_{\mathrm{f}}/v$ represents the Yukawa couplings, $v=246\GeV$ is the Higgs field vacuum expectation value, $g_{\chi}$ is the DM-mediator coupling, and $g_{\cPq}$ is the fermion-mediator coupling. The mediator particle subsequently decays into DM particles, which escape detection and leave an imbalance of momentum in the transverse plane, referred to as \ptmiss.
Several theoretical studies of these types of models have been performed, in which the third-generation quark is either a top or bottom quark, leading to the production of DM in association with a pair of top (\ttDM) or bottom (\bbDM) quarks, respectively~\cite{PhysRevD.88.063510,PhysRevD.91.015017,Haisch:2015ioa,Haisch:2012kf}. The main production diagram for \ttDM processes is shown in Fig.~\ref{fig:STdiagram} (upper left).

Previous searches in these final states have been carried out by the ATLAS and CMS Collaborations at center-of-mass energies of 8\TeV \cite{Khachatryan:2015nua,Aad:2014vea} and 13\TeV \cite{EXO16005paper,Aaboud:2017rzf,Sirunyan:2018dub}.
While the former results are based on an effective field theory (EFT) approach, the latter ones are interpreted
 in the context of simplified DM scenarios, where the mediator particle is explicitly modeled in the interaction.
These interpretations have so far neglected the contribution from DM production in association with a single top quark (\tDM) in which the interaction is mediated by a neutral spin-0 particle, as pointed out in Ref.~\cite{PhysRevD.96.035031}.
As in the SM, the single top quark is produced through processes mediated by a virtual $t$ channel (Fig.~\ref{fig:STdiagram}, upper right) or through associated production with a {\PW} boson (Fig.~\ref{fig:STdiagram}, lower left and right)~\cite{PhysRevD.96.035031}.
While the $s$ channel production of a {\PW} boson is also possible, this process is found to have a negligible contribution for this search.
The neutral DM mediator particle is then produced either by radiation from the top quark or via top quark fusion, as described in Ref.~\cite{Haisch:2016gry} for the associated production of DM with a top quark pair.

In this search, \tDM processes mediated by a neutral spin-0 particle are investigated for the first time.
This additional production mechanism is predicted by the same interactions described in Eqs.~(\ref{eq:L_scalar}) and (\ref{eq:L_pseudoscalar}) that also predict \ttDM events. For this reason, in the presented search \tDM and \ttDM processes are both considered.
Searches for similar final states referred to as ``monotop'', which involve the production of a top quark and DM particles but without additional jets or {\PW} bosons, have been conducted by the CDF experiment~\cite{PhysRevLett.108.201802} at the Fermilab Tevatron, by the ATLAS and CMS Collaborations \cite{EurPhysJC.75.79,PhysRevLett.114.101801}
at the CERN LHC at center-of-mass energies of 8\TeV, and at 13\TeV by the CMS Collaboration~\cite{Sirunyan:2018gka}.
The underlying simplified models explored in these results, unlike the one presented in Eqs.~(\ref{eq:L_scalar}) and (\ref{eq:L_pseudoscalar}), assume either the resonant production of a +2/3 charged and colored spin-0 boson that decays into a right-handed top quark and one DM particle, or a spin-1 mediator with flavor changing neutral current interactions.
Considering these models, in addition to the DM particle, only one top quark is assumed to be produced in the final state, unlike the \tDM processes considered in this search where the top quark is produced through SM-like diagrams alongside a light quark or a {\PW} boson (Fig.~\ref{fig:STdiagram}).

In this paper we present a search for an excess of events above the SM background in the \ptmiss spectrum, as expected for the DM scenarios discussed earlier,
for events that contain exactly one lepton (electron or muon)  or zero leptons, henceforth assigned to the ``single-lepton'' (SL) region or to the ``all-hadronic'' (AH) region, respectively.
The sensitivity of this analysis is improved beyond that of previous analyses by introducing a categorization of these signatures and new discriminating variables, as discussed in more detail in Section~\ref{sec:Selection}.

\section{The CMS detector and event reconstruction}
The central feature of the CMS apparatus is a superconducting solenoid of 6\unit{m} internal diameter, providing a magnetic field of 3.8\unit{T}. Within the solenoid volume are a silicon pixel and strip tracker, a lead tungstate crystal electromagnetic calorimeter (ECAL), and a brass and scintillator hadron calorimeter (HCAL), each composed of a barrel and two endcap sections. Forward calorimeters extend the pseudorapidity ($\eta$) coverage provided by the barrel and endcap detectors. Muons are detected in gas-ionization chambers embedded in the steel flux-return yoke outside the solenoid. A more detailed description of the CMS detector, together with a definition of the coordinate system used and the relevant kinematic variables, can be found in Ref.~\cite{Chatrchyan:2008zzk}.

Events of interest are selected using a two-tiered trigger system~\cite{Khachatryan:2016bia}. The first level, composed of custom hardware processors, uses information from the calorimeters and muon detectors to select events at a rate of around 100\unit{kHz} within a time interval of less than 4\mus. The second level, known as the high-level trigger, consists of a farm of processors running a version of the full event reconstruction software optimized for fast processing, and reduces the event rate to around 1\unit{kHz} before data storage.

The particle-flow (PF) algorithm~\cite{CMS-PRF-14-001} aims to reconstruct and identify each individual particle in an event, with an optimized combination of information from the various elements of the CMS detector. The energy of photons is obtained directly from the ECAL measurement and corrected for zero-suppression effects. The energy of electrons is obtained from a combination of the electron momentum at the primary interaction vertex as determined by the tracker, the energy of the corresponding ECAL cluster, and the energy sum of all bremsstrahlung photons spatially compatible with originating from the electron track.
The muon track is obtained from the combination of central tracker and muon system information, and its curvature provides an estimate of the momentum.
The energy of charged hadrons is determined from a combination of their momentum measured in the tracker and the matching ECAL and HCAL energy deposits, corrected for zero-suppression effects and for the response function of the calorimeters to hadronic showers. Finally, the energy of neutral hadrons is obtained from the corresponding corrected ECAL and HCAL energy.

The reconstructed vertex with the largest value of summed physics-object $\pt^2$, where \pt is the transverse momentum, is taken to be the primary proton-proton (\Pp\Pp) interaction vertex. The physics objects are the jets and the associated \ptvecmiss, taken as the negative vector \pt sum of those jets.
For each event, hadronic jets are clustered from the particles reconstructed with PF (PF candidates) using the infrared- and collinear-safe anti-\kt algorithm~\cite{Cacciari:2008gp, Cacciari:2011ma} with a distance parameter of 0.4. The jet momentum is determined as the vectorial sum of all particle momenta in the jet, and is found from simulation to be within 5 to 10\% of the parton's generated momentum over the whole \pt spectrum and detector acceptance. Additional {\Pp\Pp} interactions within the same or nearby bunch crossings (pileup) can contribute additional tracks and calorimetric energy depositions to the jet momentum. To mitigate this effect, tracks identified as originating from pileup vertices are discarded and an offset correction is applied to correct for remaining contributions~\cite{Cacciari:2008gn}.
Jet energy corrections are derived from simulation and applied to calibrate the jet momentum. In situ measurements of the momentum balance in dijet, photon+jet, \zjets, and multijet events are used to account for any residual differences in jet energy scale in data and simulation~\cite{CMS-PAS-JME-16-003}.
Additional selection criteria are applied to each jet to remove jets potentially dominated by anomalous contributions from various subdetector components or reconstruction failures~\cite{CMS-PAS-JME-16-003}.

The combined secondary vertex {\cPqb} tagging algorithm (CSVv2) is used to identify jets originating from the hadronization of bottom quarks~\cite{Sirunyan:2017ezt}, denoted in the following as ``{\cPqb}-tagged jets''.
At the operating point of the tagging algorithm chosen for this analysis, the efficiency of identifying {\cPqb} quark jets in simulated \ttbar events is about 80\%, integrated over \pt, and the misidentification rate for light-flavor jets is about 1\%.
Scale factors are applied to the simulated samples in order to reproduce the {\cPqb} tagging performance measured in data.

The missing transverse momentum vector \ptvecmiss is defined as the negative vector \pt sum of all PF particles originating from the primary vertex; its magnitude is defined as \ptmiss. Jet energy scale and resolution corrections are also propagated to the \ptvecmiss calculation.

\section{Data sample and simulation}

The data used in this search were recorded with the CMS detector in 2016 and correspond to an integrated luminosity of 35.9\fbinv. Several trigger criteria were used to collect the data, either requiring large amounts of \ptmiss or the presence of at least one high-\pt lepton (electron or muon).
Simulated samples are corrected to reproduce the observed trigger efficiencies in data.

Specifically, events that do not contain leptons are selected if they have
\ptmiss and missing hadronic activity \mht~\cite{Khachatryan:2016bia} above 120\GeV.
This trigger is nearly 100\% efficient for events with \ptmiss of at least 250\GeV.
The second set of triggers requires the presence of at least one isolated electron (muon) with $\pt>27$ (25)\GeV.
The corresponding trigger efficiencies are above 90\% for leptons with $\pt>30\GeV$.
Trigger efficiencies are measured in data.

Monte Carlo (MC) simulated samples of the main SM backgrounds and of the DM signal processes are used to optimize the event selection, assess our sensitivity to the new-physics scenarios, and form the basis of our background estimation strategy.
While the detailed background composition depends on the specific channel, the main sources arise from \ttjets, \wjets, and \zjets production.
Simulated events of  \ttjets production and single top quark processes are generated at next-to-leading order (NLO) in quantum chromodynamics (QCD) using {\POWHEG v2} and {\POWHEG v1}~\cite{Nason:2004rx,Frixione:2007vw,Alioli:2010xd}, respectively. For \ttjets processes, the top quark \pt distribution is reweighted to reproduce the differential cross section obtained from CMS measurements~\cite{Khachatryan:2016mnb}.
Samples of \zjets, \wjets, and QCD multijet events are generated at leading order (LO) using \MGvATNLO~\cite{Alwall:2014hca} with the MLM prescription~\cite{Mangano:2006rw} for matching jets from the matrix element (ME) calculation to the parton shower description.
Dedicated electroweak~\cite{Denner:2009gj, Denner:2011vu, Denner:2012ts, Kuhn:2005gv, Kallweit:2014xda, Kallweit:2015dum} and QCD (calculated with \MGvATNLO) NLO/LO $K$ factors, parametrized  as functions of the generated boson \pt, are applied to \zjets and \wjets events.
Other SM backgrounds include rare processes, such as \ensuremath{\ttbar}+{\PW} and \ensuremath{\ttbar}+\PZ, which are simulated based on the NLO ME calculations implemented in \MGvATNLO~and the FxFx~\cite{Frederix:2012ps} prescription to merge multileg processes. Diboson processes ($\PW\PW$, $\PW\PZ$, $\PZ\PZ$, $\PW\PH$, $\PZ\PH$) are generated at NLO using either \MGvATNLO~or {\POWHEG v2}.
All background samples are normalized using the most accurate cross section calculations available, which generally incorporate NLO or next-to-NLO (NNLO) precision.

The signal process is simulated at LO with the \MGvATNLO~v2.4.2 event generator using a simplified model investigated within the LHC Dark Matter Forum~\cite{Abercrombie:2015wmb}. In this model, the DM particles $\chi$ are assumed to be Dirac fermions and the mediators are spin-0 particles $\phi$ (\Pa) that couple preferentially to third-generation SM quarks through scalar (pseudoscalar) couplings whose strengths are parametrized by the factor $g_{\cPq}$. The coupling strength between the mediator and the DM particles is in turn given by the factor $g_{\chi}$.
This simplified model has a minimal set of four free parameters:
($\mchi,\ m_{\phi/\Pa},\ g_{\chi},\ g_{\cPq}$), and the benchmark scenarios assume $g_{\chi}=g_{\cPq}=1$ as per recommendations of the LHC Dark Matter Working Group~\cite{Boveia:2016mrp}. In addition, in this search we focus on the $\mchi=1\GeV$ benchmark, which is a convenient signal reference as the production cross section is almost independent of \mchi for on-shell mediators~\cite{Abercrombie:2015wmb}.
  This simplified spin-0 model does not account for mixing between the $\phi$ scalar mediator and the SM Higgs boson, as discussed in Ref.~\cite{Bauer:2016gys}.
Under these assumptions two distinct DM scenarios are possible: the associated production with a top quark pair (\ttDM) and the associated production with a single top quark (\tDM).
Cross sections for both signal processes are calculated at LO with \MGvATNLO~v2.4.2, with one (zero) additional partons for \ttDM (\tDM) events.

For all simulated samples, the initial-state partons are modeled with the NNPDF 3.0~\cite{Ball:2014uwa} parton distribution function (PDF) sets at LO or NLO in QCD to match the ME calculation.
Generated events are interfaced with {\PYTHIA 8.205}~\cite{Sjostrand:2007gs} for parton showering and hadronization using the \textsc{CUETP8M1} tune~\cite{Khachatryan:2015pea}, except for simulated \ttjets events where the \textsc{CUETP8M2} tune customized by CMS with an updated strong coupling \alpS for initial-state radiation is employed~\cite{CMS-PAS-TOP-16-021}.
 All signal and background samples are processed using \GEANTfour~\cite{Agostinelli:2002hh} to provide a full simulation of the CMS detector, including a simulation of the previously mentioned
 triggers. Correction factors are derived and applied to the simulated samples to match the trigger efficiencies measured in data. Additional corrections are applied to cover
remaining residual differences between data and simulation that arise from the lepton identification and reconstruction efficiencies, as well as from {\cPqb}-tagged jet identification efficiencies.

\section{Event selection}
\label{sec:Selection}
This search, similarly to a previous search for \ttDM events~\cite{Sirunyan:2018dub}, defines several orthogonal signal regions (SRs) that are statistically combined in a simultaneous global fit of the \ptmiss spectrum. At the same time, various improvements are incorporated into this search to enhance the sensitivity to the \tDM final state over that of previous analyses~\cite{Sirunyan:2018dub}.

At the analysis level, jet candidates are required to have $\pt>30\GeV$ and are categorized as ``central'' if they lie within $\abs{\eta}<2.4$ and as ``forward'' if they are within $2.4<\abs{\eta}<4.0$. The {\cPqb}-tagged jets identified by the CSVv2 algorithm are also required to have $\pt>30\GeV$ and in addition to lie within $\abs{\eta}<2.4$.
Electrons and muons are selected with $\pt>30\GeV$ and $\abs{\eta}<2.1$. Events containing additional leptons with $\pt>10\GeV$ and $\abs{\eta}<2.1$ are vetoed.
To ensure that candidate leptons are well-measured, identification requirements, based on hit information in the tracker and muon systems and on energy deposits in the calorimeters, are imposed.
Leptons are further required to be isolated from hadronic activity, to reject leptons within jets that could arise, for example, from the decay of {\cPqb} quarks. A relative isolation quantity is defined as the scalar \pt sum of all PF candidates within a $\Delta{R}=\sqrt{\smash[b]{(\Delta\eta)^{2}+(\Delta\phi)^{2}}}$ cone of radius 0.3 (0.4) centered around the electron (muon) candidate, where $\phi$ is the azimuthal angle in radians, divided by the lepton \pt~\cite{CMS-PAPER-MUO-10-004,Khachatryan:2015hwa}.
This relative isolation is required to be less than 0.059\,(0.057) for electrons in the barrel (endcap) and less than 0.15 for muons.

Events are separated into orthogonal categories based on the number of {\cPqb}-tagged jets (\nbjet), with $\nbjet=1$ or $\nbjet\geq2$, and additional requirements on the number of forward jets are placed (0 or ${\geq}1$ forward jets) for the $\nbjet=1$ category.
The mentioned categorization in terms of forward jets allows a further enhancement of \tDM $t$ channel events. In fact, as shown in Fig. 1, this production mode leads to final states with one top quark and an additional jet, which tends to be in the forward region of the detector, while the additionally produced {\cPqb} quark is typically low in \pt and therefore is not reconstructed.
The minimum requirements on the number of jets is also lowered, with respect to the previous searches, to enhance the sensitivity specifically to the \tDM model.
Control regions (CRs) enriched in the major background processes are included in the fit in order to improve the estimates of the background contributions.

Events are classified into two ``channels'', based on the number of leptons in the final state from the top quark decay: the single-lepton SL channel, containing events with exactly one electron or muon with $\pt>30\GeV$, and the all-hadronic AH channel, containing events with exactly zero leptons with $\pt>10\GeV$.
A set of discriminating variables is identified, as discussed in more detail in Sections~\ref{sec:Selection_SL} and~\ref{sec:Selection_AH} for the SL SRs and the AH SRs, respectively. The selection requirements on these variables are optimized simultaneously to increase the signal significance, using as a figure of merit the ratio between the expected number of signal and the square root of the expected SM background events.
The considered signal events are either \tDM events for a region that contains exactly one {\cPqb}-tagged jet ($\nbjet=1$) or \ttDM events for a region that contains two or more {\cPqb}-tagged jets ($\nbjet\geq2$).
The region with exactly one {\cPqb}-tagged jet is further divided into exactly zero or ${\geq}1$ forward jets.

\subsection{Single-lepton signal regions}
\label{sec:Selection_SL}
Events in the SL channel are required to contain ${\geq}1$ identified {\cPqb}-tagged jet, at least 2 jets with $\pt>30\GeV$, and $\ptmiss>160\GeV$.
After this selection, the dominant backgrounds in the SL channel are from \ttbar and \wjets processes.
Other backgrounds include single top quark, Drell--Yan, and diboson production.

To further improve the sensitivity and to reduce the dominant background from single-lepton \ttbar and \wjets processes, we impose a requirement on the transverse mass \mT, calculated as:
\begin{equation}
\mT = \sqrt{\smash[b]{2\ptmiss \pt^{\ell}[1-\cos(\Delta\phi)]}},
\label{eq:mTW}
\end{equation}
where $\pt^{\ell}$ is the transverse momentum of the lepton and $\Delta\phi$
is the opening angle between the lepton direction and the \ptmiss vector in the transverse plane.
The \mT variable is constrained by kinematic properties to be less than the {\PW} boson mass
for leptonic on-shell {\PW} decays in \ttbar and \wjets events, while for signals, off-shell {\PW} decays, or for dileptonic decays of \ttbar, the \mT variable is expected to exceed the {\PW} mass because of the additional \ptmiss in the event. A requirement of $\mT>160\GeV$ therefore reduces the background from single-lepton events significantly and enhances the analysis sensitivity to the DM models.

After the \mT selection, the remaining \ttbar background is primarily from events where both top quarks decay leptonically (\tttwol) and one lepton is not identified.
This background can be further reduced by making use of the \mTt variable~\cite{Bai:2012gs}, which is defined as the minimal value of the mass of a particle assumed to be pair produced and to decay to a {\PW} boson and a {\cPqb} quark jet. The {\PW} bosons are assumed to be produced on-shell and to decay leptonically, where one of the two leptons is not detected. Based on the variable definition, in \tttwol
events the \mTt distribution has a kinematic end point at the top quark mass,
assuming perfect detector response, while this is not the case for signal events
where two additional DM particles are present.
The calculation of \mTt requires two {\cPqb}-tagged jets from the decay of the top quarks, where one of these {\cPqb}-tagged jets comes from the same decay chain as the reconstructed lepton.
If only one {\cPqb}-tagged jet is identified in the event, each of the first three (or two in three-jet events)
leading non-{\cPqb}-tagged jets is considered as the second {\cPqb}-tagged jet in the calculation.
The \mTt is then evaluated for all possible jet-lepton combinations
and the minimum \mTt value is considered to discriminate between signal and background events.
If two or more {\cPqb}-tagged jets are identified in the events, all {\cPqb}-tagged jets are considered
and similarly all possible jet-lepton combinations are used to calculate \mTt values.
The smallest of all the \mTt values is taken as the event discriminant.

In addition, jets and the \ptvecmiss vector tend to be more separated in the transverse plane in signal
events than in \ttbar background processes. To improve the search sensitivity, the minimum opening angle \mindphi in the transverse plane between the direction of each of the first two leading-\pt jets with $\abs{\eta}<2.4$ and the \ptvecmiss vector is required to be greater than $1.2$ radians.

The \ttbar background is further reduced by requiring that the transverse mass \mTb of the \ptvecmiss vector and of a {\cPqb}-tagged jet is greater than 180\GeV, where \mTb is defined similarly to Eq.~(\ref{eq:mTW}) but considering a \cPqb-tagged jet instead of a lepton.
In fact, for the remaining \ttbar background \mTb tends to have values below or around the top quark mass if the {\cPqb}-tagged jet belongs to the top quark whose lepton is not identified.
For the calculation we choose the {\cPqb}-tagged jet with the highest CSVv2 discriminant value, if there is more than one candidate.

A summary of the selection criteria for the SL SRs is shown in the first three columns of Table~\ref{tab:SR}. Each region is identified by a unique name, where 0$\ell$ denotes exactly zero leptons, 1(2) \cPqb-tag represents exactly 1 (at least 2) \cPqb-tagged jet, and 0 FJ or 1 FJ denotes exactly zero or at least one forward jet.

\begin{table}[htbp]
\centering
\topcaption{Final event selections for the SL and AH SRs. Electrons and muons are kept separate for the SL channel.}
\cmsTable{
\begin{tabular}{lccc{c}@{\hspace*{5pt}}ccc}
\hline
& \multicolumn{ 3}{c}{Single-lepton SRs} &  & \multicolumn{ 3}{c}{All-hadronic SRs} \\\cline{2-4}\cline{6-8}
& 1$\ell$, 1 \cPqb-tag, 0 FJ & 1$\ell$, 1 \cPqb-tag, 1FJ & 1$\ell$, 2 \cPqb-tag & & 0$\ell$, 1 \cPqb-tag, 0 FJ & 0$\ell$,1 \cPqb-tag, 1 FJ & 0$\ell$, 2 \cPqb-tag\\
\hline
Forward jets & $=$0  & ${\geq}1$ & \NA       &   & $=$ 0 & ${\geq}1$ & \NA \\
\nbjet       & $=$1  & $=$1      & ${\geq}2$ &   & $=$ 1 & $=$1      & ${\geq}2$ \\
\nlep        & $=$1  & $=$1      & $=$1      &   & $=$ 0 & $=$0      & $=$0 \\[\cmsTabSkip]
\HTratio     & \multicolumn{ 3}{c}{\NA}               & &   \multicolumn{ 2}{c}{\NA}       & $<$0.5 \\
\njet        & \multicolumn{ 3}{c}{${\geq}2$}         & &  \multicolumn{ 3}{c}{${\geq}3$}          \\
\ptmiss      & \multicolumn{ 3}{c}{$>$160\GeV}        & &  \multicolumn{ 3}{c}{$>$250\GeV}         \\
\mT          & \multicolumn{ 3}{c}{$>$160\GeV}        & &  \multicolumn{ 3}{c}{\NA}                \\
\mTt         & \multicolumn{ 3}{c}{$>$200\GeV}        & &  \multicolumn{ 3}{c}{\NA}                \\
\mindphi     & \multicolumn{ 3}{c}{$>$1.2\unit{rad.}} & &  \multicolumn{ 3}{c}{$>$1.0\unit{rad.}}  \\
\mTb         & \multicolumn{ 3}{c}{$>$180\GeV}        & &  \multicolumn{ 3}{c}{$>$180\GeV}         \\
\hline
\end{tabular}
}
\label{tab:SR}
\end{table}

\subsection{All-hadronic signal regions}
\label{sec:Selection_AH}
Events categorized into the AH channel must contain at least 1 identified {\cPqb}-tagged jet and at least 3 jets with $\pt>30\GeV$, $\ptmiss>250\GeV$, and \mindphi greater than 0.4 radians.

The dominant backgrounds after this selection arise from \ttbar, \wjets, and \Zinv processes. Other backgrounds include QCD multijet events, single top quark, Drell--Yan, and diboson production.

Semileptonic \ttbar events populate this channel if the lepton in the final state is not identified. This \ttonel background is reduced by applying the same \mTb selection as introduced in the SL channel.
To further reduce the \ttonel  background, together with that from \Zinv events, we make use of the \HTratio variable, which is defined as the ratio of the leading \pt jet in the event divided by the total hadronic transverse energy in the event, \HT, which is the scalar
 \pt sum of the jets with $\pt>30\GeV$ within $\abs{\eta}<2.4$.
In the case of background, the distribution peaks at higher values with respect to \ttDM signal events.
The   \tDM events, however, tend to exhibit a distribution similar to that of the background.
Events in the $\nbjet\geq2$ category are required to have $\HTratio<0.5$.

For QCD multijet events no intrinsic \ptmiss is expected. Therefore, events that pass our minimum \ptmiss selection contain mostly \ptmiss which arises from jet mismeasurements. For these events, the \ptmiss is often aligned with one of the leading jets. As a result, selecting events with \mindphi values greater than 1 radian reduces the background from QCD multijet production. This contribution to the SR, estimated through simulated samples, is negligible. The description of the QCD multijet background basic kinematic distributions is verified in a dedicated region enriched in multijet events, obtained by reversing the \mindphi selection, and the simulation is found to model the data well.

A summary of the selection criteria for the AH SRs is shown in the last three columns of Table~\ref{tab:SR}. Each region is identified by a unique name, where 1$\ell$ denotes exactly one muon or one electron, 1(2) \cPqb-tag represents exactly 1 (at least 2) \cPqb-tagged jet, and 0 FJ or 1 FJ denotes exactly zero or at least one forward jet.

\subsection{Control regions}

After events are categorized according to the selection presented in Table~\ref{tab:SR}, the expected SM backgrounds in these different regions must be evaluated.
In the SL SRs, the main backgrounds are dileptonic \ttbar events, where one lepton is not identified, and \wjets events.  For the AH regions the main backgrounds arise instead from single-lepton \ttbar and \wjets events, where the lepton is not identified, and {\cPZ} boson
production, where the {\cPZ} boson decays into two neutrinos and leads to a background with genuine \ptmiss.

In order to improve the estimation of these main backgrounds, methods based on control samples in data are used. In particular, CRs enhanced in the different background sources are used to derive correction factors as a function of the \ptmiss from the comparison of the \ptmiss distribution between the data and the simulation. These corrections are extracted and simultaneously propagated across the CRs and SRs for a given channel in the context of a global fit, as explained in more detail in Section~\ref{sec:fit}. The residual backgrounds processes are modeled with simulation.

The background CRs for the SL and AH channels are designed to be statistically independent from the corresponding SRs.

\subsubsection{Single-lepton control regions}
\label{sec:SLCR}
The first set of CRs is defined to isolate dileptonic \ttbar events by requiring exactly two leptons (1 electron and 1 muon, 2 electrons, or 2 muons), $\njet\geq2$, $\nbjet\geq1$, and $\ptmiss>160\GeV$.
In order to statistically enhance  these CRs the \mT, \mTt, and forward jet selections are removed.

The second set of CRs is designed to isolate \wjets events by requiring exactly one lepton (electron or muon), $\njet\geq2$, $\nbjet=0$,  $\ptmiss>160\GeV$, and $\mT>160\GeV$. The $\nbjet=0$ requirement makes this CR orthogonal to the SL SR and allows the events in the \mT tail to be modeled without extrapolation from a lower-\mT region.

Both of these selections are summarized in the first two columns of Table~\ref{tab:CR}.

\begin{table}[htbp]
\centering
\topcaption{Control regions defined for the main backgrounds of the SL SRs (first two columns, \tttwol and \wjets) and the AH SRs (last 3 columns, \ttonel, \wjets, and \ZLL). Some selections applied in the SRs are removed in the corresponding CRs to increase the available statistics and are therefore not listed. The \ptmiss selection for the \ZLL CR refers to the hadronic recoil.}
\begin{tabular}{lcc{c}@{\hspace*{5pt}}ccc}
\hline
 & \multicolumn{2}{c}{Single-lepton CRs} & & \multicolumn{3}{c}{All-hadronic CRs} \\ \cline{2-3}\cline{5-7}
 & CR \tttwol & CR {\PW}($\ell \nu$)     & & CR \ttonel &CR {\PW}($\ell \nu$) &CR {\cPZ}($\ell \ell$) \\
\hline
\nbjet   & ${\geq}1$  & $=$0       & & ${\geq}1$         & $=$0       & $=$0 \\
\nlep    & $=$2       & $=$1       & & $=$1              & $=$1       & $=$2 \\
\njet    & ${\geq}2$  & ${\geq}2$  & & ${\geq}3$         & ${\geq}3$  & ${\geq}3$  \\
\ptmiss  & $>$160\GeV & $>$160\GeV & & $>$250\GeV        & $>$250\GeV & $>$250\GeV\\
\mT      & \NA        & $>$160\GeV & & $<$160\GeV        & $<$160\GeV & \NA \\
\mindphi & \NA        & \NA        & & $>$1.0\unit{rad.} & \NA        & \NA \\
\mll     & \NA        & \NA        & & \NA               & \NA        & $[60,120]$\GeV \\
\hline
\end{tabular}
\label{tab:CR}
\end{table}

\subsubsection{All-hadronic control regions}

For the AH SRs, three independent sets of CRs are defined.
The first set of CRs is enhanced in single-lepton \ttbar events selecting events
 with exactly one lepton (electron or muon), $\njet\geq3$,  $\nbjet\geq1$,  $\ptmiss>250\GeV$, and, in order to avoid overlap with the SL SRs, $\mT<160\GeV$.

The second set of CRs is defined to enhance single-lepton \wjets events.
Events are selected with exactly one lepton (electron or muon), $\njet\geq3$,  $\nbjet=0$, $\ptmiss>250\GeV$, and in order to avoid overlap with the SL \wjets CR, $\mT<160\GeV$.

The third and last set of CRs are designed to model the background due to \zjets production, where the {\PZ} boson decays into a pair of neutrinos (\Zinv).
Here we use the {\PZ} boson decays to an opposite-sign, same-flavor dilepton pair (\ZLL), as proxy events to emulate the kinematic properties of the \zjets process.
Events are selected requiring 2 leptons, which have the same flavor (\ie, {\Pe\Pe} or $\mu\mu$), and opposite charge, and that satisfy a requirement on their invariant mass of $60<\mll<120\GeV$.
Additionally, events must contain at least 3 jets, but events with {\cPqb}-tagged jets are vetoed ($\nbjet=0$).
In order to reproduce the \pt spectrum of \Zinv events, the two leptons are added to the
\ptvecmiss, referred to as hadronic recoil.

A summary of the different AH CRs can be found in the last three columns of Table~\ref{tab:CR}.

\section{Systematic uncertainties}
\label{sec:systematics}
Several sources of uncertainty are considered that affect either the simulation of the background processes or the underlying theoretical modeling.
We distinguish between two types of uncertainties, ones that only affect the normalization of a process and others that additionally affect the shape of the \ptmiss distribution.
All of these uncertainties are included in the global simultaneous fit, described in detail later.
The largest impacts on the final results stem from the uncertainties in the {\cPqb} tagging scale factors and the limited statistical precision of the dilepton \ttbar CR, where the latter
is the main determining factor for the contribution of \ttbar events in the SL SRs.

The following sources of uncertainty correspond to constrained normalization nuisance parameters in the fit (unless specified, the source of uncertainty applies to all search channels):

\begin{itemize}
\item \textit{Lepton reconstruction, selection, and trigger}. Scale factors are applied to the simulation in order to mimic the measured lepton reconstruction and selection efficiencies in data.
 The measured uncertainties in these scale factors are of the order of 2.2\% per electron and 1\% per muon, and are \pt and $\eta$ dependent~\cite{CMS-PAPER-MUO-10-004,Khachatryan:2015hwa}. The effect of these uncertainties is found to be independent of the \ptmiss spectrum.

\item \textit{\ptmiss trigger}. At values of $\ptmiss>250\GeV$ the applied triggers are almost fully efficient; a normalization uncertainty of 2\% is assigned. This uncertainty is only applied in the AH channel.

\item \textit{{\cPqb} tagging efficiency scale factors}. The {\cPqb} tagging and light-flavor mistag efficiencies scale factors and the respective uncertainties are measured in independent control samples~\cite{Sirunyan:2017ezt}, and propagated to the analysis. In the range of \ptmiss considered, these scale factor uncertainties do not alter the shape of the \ptmiss distribution.

\item \textit{Forward jets}. Inclusive CRs in terms of forward jet multiplicity are considered to constrain the major background in the 0 and ${\geq}1$ forward jets SRs. The impact of this extrapolation in forward jets multiplicity on the background estimation is evaluated and assigned as additional systematic uncertainty. The extrapolation effect is evaluated by splitting each CR into a 0f and 1f category, and a systematic uncertainty is assigned based on the ratio of the correction factors, where each correction factor is the ratio of the data to the simulation in its category. This uncertainty ranges from approximately 2\% (\wjets AH) to about 7\% (\ttbar SL).

\item \textit{Pileup modeling}.
Systematic uncertainties due to pileup modeling are taken into account by varying the total inelastic cross section used to calculate pileup distributions in simulation by $\pm$4.6\%~\cite{Sirunyan:2018nqx}.

\item \textit{Luminosity}. An uncertainty of 2.5\% is taken on the integrated luminosity of the data sample~\cite{CMS-PAS-LUM-17-001}.

\item \textit{QCD multijet background normalization}. An uncertainty of 100\% in the normalization is considered for QCD processes to cover effects in the kinematic tails that may not be well-modeled by the simulation.
This has little overall impact on the final result, since the contribution from QCD multijet events is reduced to a negligible amount in this analysis.

\item \textit{Single top quark background normalization}.
An uncertainty of 20\% in the normalization is considered for single top quark processes, accounting for the uncertainty in the PDF and the effects from varying the factorization and renormalization scale parameters.

\item \textit{Uncertainty related to ECAL mistiming}.
Partial mistiming of signals in the forward regions of the ECAL endcaps led to a minor reduction in trigger efficiency.
To cover this effect, an additional uncertainty is applied on the signal acceptances of up to 10\% in the forward jet categories.
A potential effect on the background extrapolation into regions with forward jets is already taken into account by a dedicated systematic uncertainty.

\end{itemize}

The following sources of uncertainty affect the shape of the \ptmiss distribution, as well as the normalization of the various backgrounds and the signal, and are applied to all search channels:

\begin{itemize}

\item \textit{Jet energy scale}. Reconstructed jet four-momenta in the simulation are varied according to the uncertainty in the jet energy scale. Jet energy scale uncertainties are coherently propagated to all observables, including \ptmiss~\cite{Khachatryan:2016kdb}.

\item \textit{PDF uncertainties}. Uncertainties due to the choice of PDF are estimated by reweighting the samples with the NNPDF3.0~\cite{Ball:2014uwa} replicas~\cite{Butterworth:2015oua} and are applied to all backgrounds except for the single top quark, as these uncertainties are covered by the associated background normalization uncertainty.

\item \textit{{\PW}/{\PZ}+heavy-flavor fraction}. The uncertainty in the fraction of  {\PW}/{\PZ}+heavy-flavor (HF) jets in \wjets and \zjets event is taken into account. The relative contribution of {\PW}+HF and {\PZ}+HF are allowed to vary  within 20\%~\cite{PhysRevD.95.052002,Khachatryan2017,CMS-PAS-SMP-12-017,Khachatryan:2016iob}.

\item \textit{Electroweak and QCD $K$ factors}. Uncertainties in the NLO/LO $K$ factors calculated for \wjets and \zjets processes are considered.  These uncertainties account for missing higher-order corrections.  For QCD, this comes from variations due to factorization and renormalization scales.  For electroweak processes, an estimate of the size of the missing higher-order corrections is obtained by taking the difference between applying and not applying the NLO/LO electroweak $K$ factors.

\item \textit{Top quark \pt reweighting}. Differential measurements of the top quark \pt spectrum in top quark pair production events~\cite{Khachatryan:2016mnb} show that the measured \pt spectrum is softer than in simulation.
In order to improve the description of top quark pair events, simulated samples are reweighted to match the measurements.
An associated systematic uncertainty is estimated by taking the difference between applying and not applying the reweighting.

\item \textit{Factorization and renormalization scales}.
 The uncertainties in the choice of the factorization and renormalization scale parameters
 are taken into account for the \ttbar, \ensuremath{\ttbar}+V, and diboson processes by applying a set of weights that represent a change of these scales by a factor of 2 or 0.5.

\item \textit{Simulation sample size}. Uncertainties due to the limited size of the simulated signal and background samples are included by allowing each bin of the distributions used in the signal extraction to fluctuate independently according to the statistical uncertainties in simulation, following Ref.~\cite{BARLOW1993219}.

\end{itemize}

\section{Signal extraction}
\label{sec:fit}

As previously discussed, the potential DM signal is expected to have the signature of \ttbar or single top quark events with additional \ptmiss, therefore leading to an excess of events above the SM prediction in the \ptmiss spectrum.
The DM signal is extracted from a simultaneous fit to the binned \ptmiss distribution in the various SRs and CRs, including all previously mentioned uncertainties. This global fit is performed as a binned maximum likelihood fit employing the \textsc{RooStats} statistical package~\cite{Moneta:2010pm}.
The main SM backgrounds were discussed previously in Section~\ref{sec:Selection}, and are dileptonic \ttjets and \wjets events for the SL SRs, and \Zinv, single-lepton \ttjets, and \wjets events for the AH SRs.

The effect of the systematic uncertainties in the shape and normalization of the \ptmiss spectrum, as discussed in the previous section,
is taken into account by introducing nuisance parameters, which are constrained by the magnitude of the corresponding source of uncertainty.
Uncertainties that affect normalization only are modeled using nuisances with log-normal probability densities.
These parameters are treated as correlated between \ptmiss bins and between the different CRs and SRs within each channel. The sources common between SL and AH SRs and CRs are correlated across channels.

To improve the estimation of the main backgrounds, an unconstrained multiplicative parameter is assigned separately to each background for each bin of the \ptmiss spectrum. These multiplicative parameters scale the normalization of the associated background process simultaneously in the SRs and CRs for a given channel.
For example, in a given \ptmiss bin of the SL selection, there is one multiplicative parameter for \ttbar that links the \ttbar background in the \ttbar enhanced 2$\ell$ CR, the \wjets enhanced 1$\ell$ CR, and the SR. Therefore, the effect of contributions of the same background process in the different CRs is also taken into account. Additionally, potential contributions from the DM signals are included for all CRs and SRs, and scaled by a signal strength modifier $\mu=\sigma/\sigma_{\text{th}}$, \ie, the ratio between the measured and theoretical cross sections. Regions containing leptons (electrons and muons) are separated by lepton flavor.

\begin{figure}[htbp!]
\centering
\includegraphics[width=0.42\textwidth]{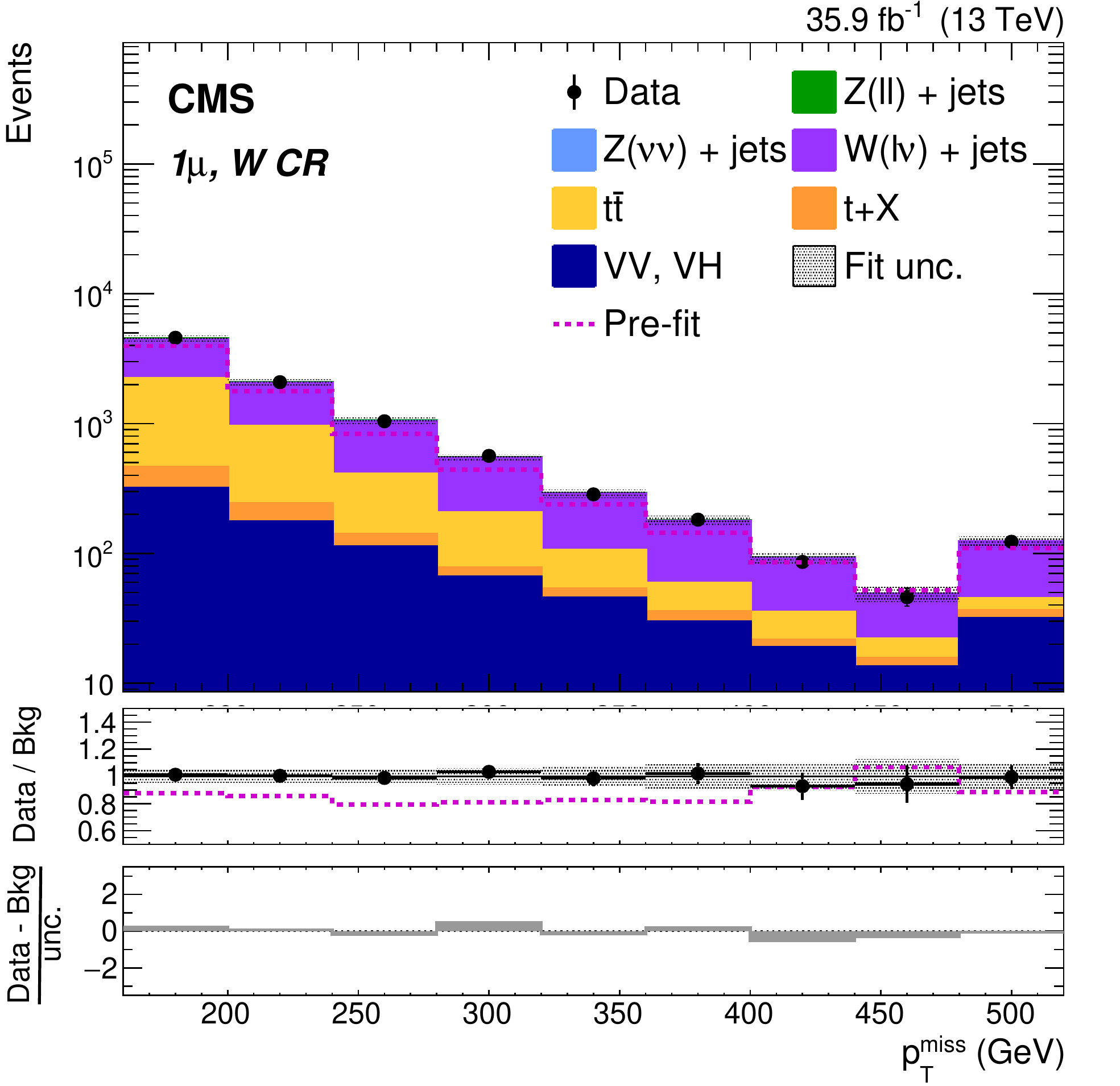}
\includegraphics[width=0.42\textwidth]{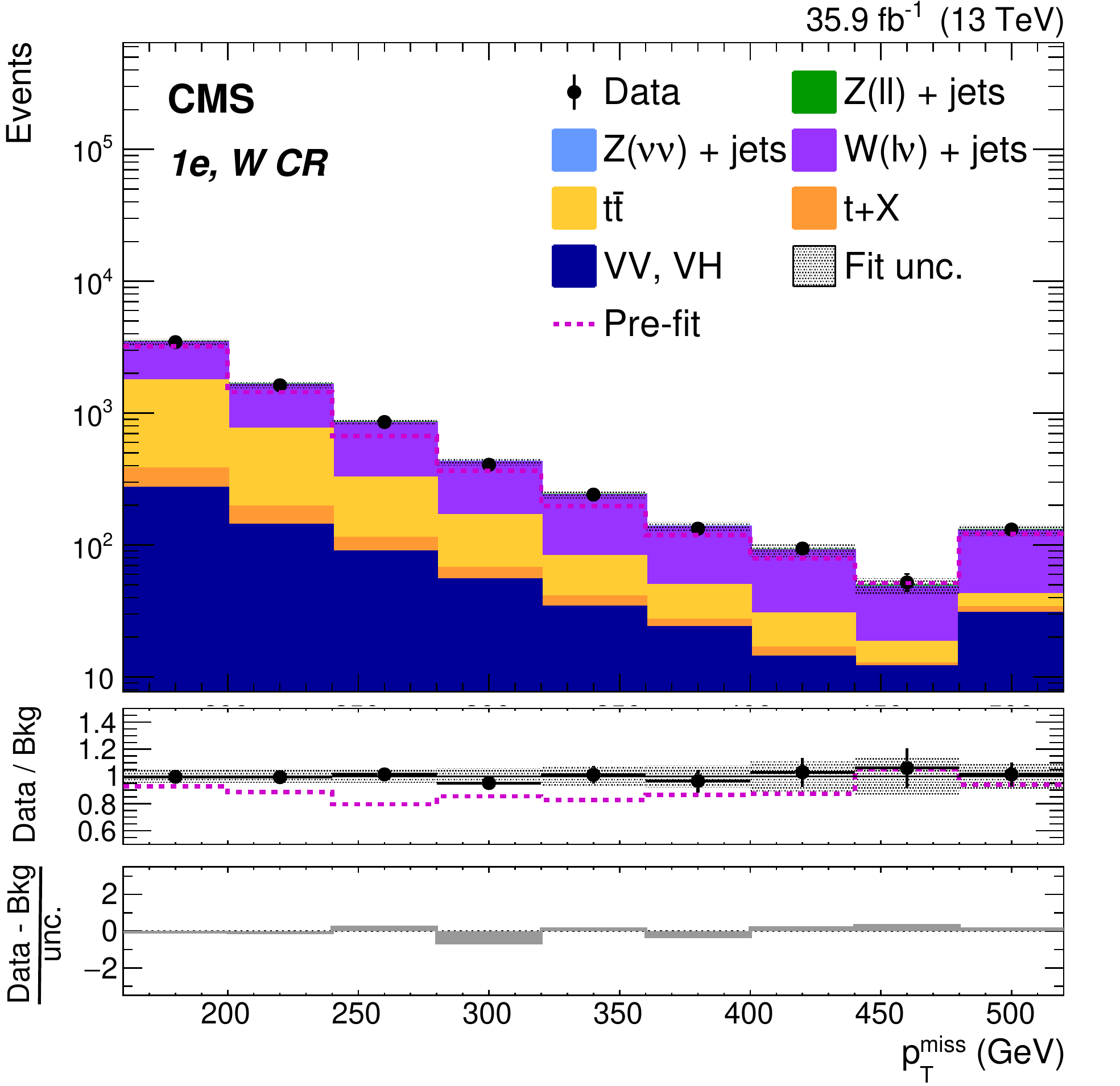}\\
\includegraphics[width=0.42\textwidth]{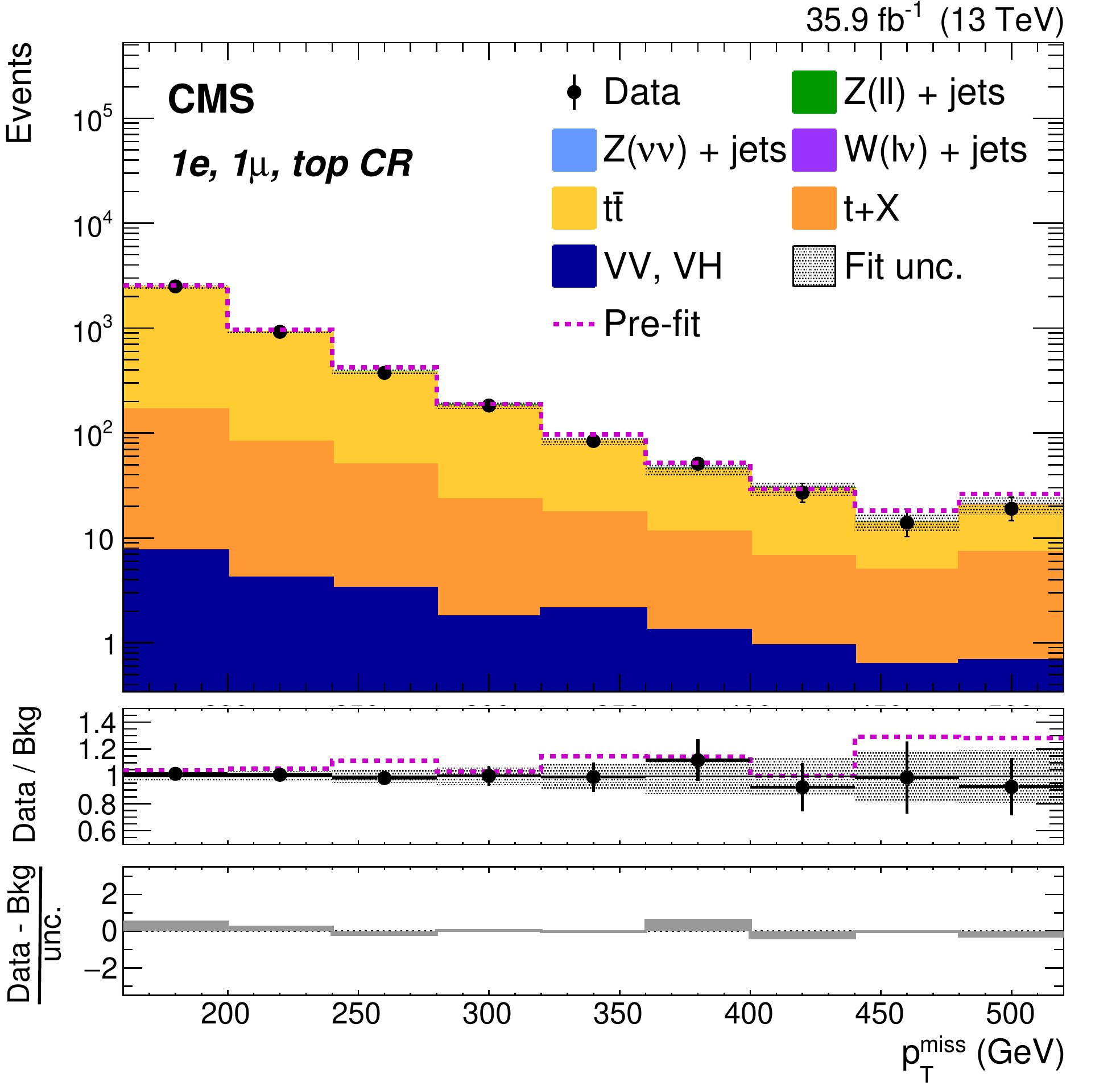}
\includegraphics[width=0.42\textwidth]{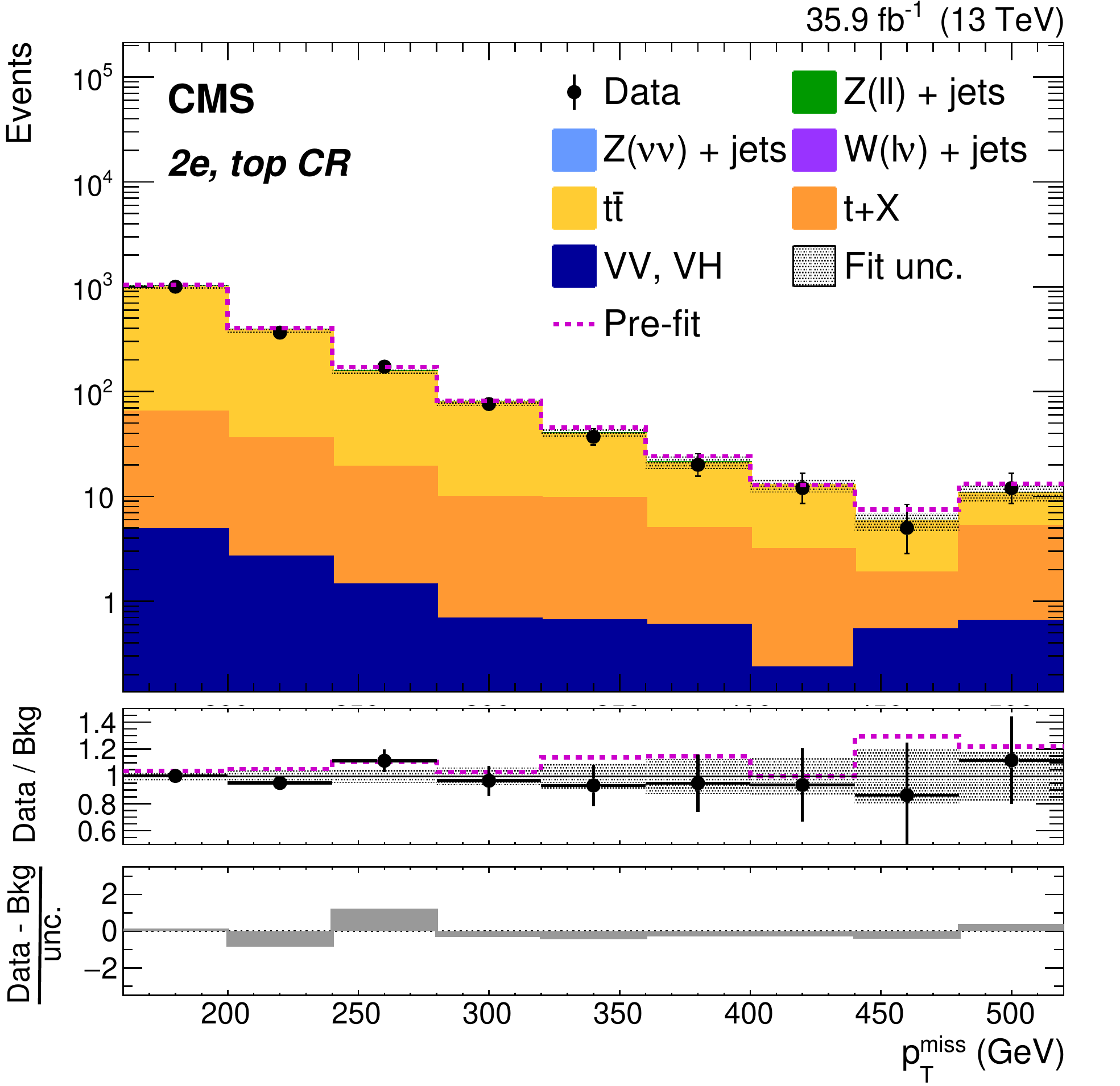}\\
\includegraphics[width=0.42\textwidth]{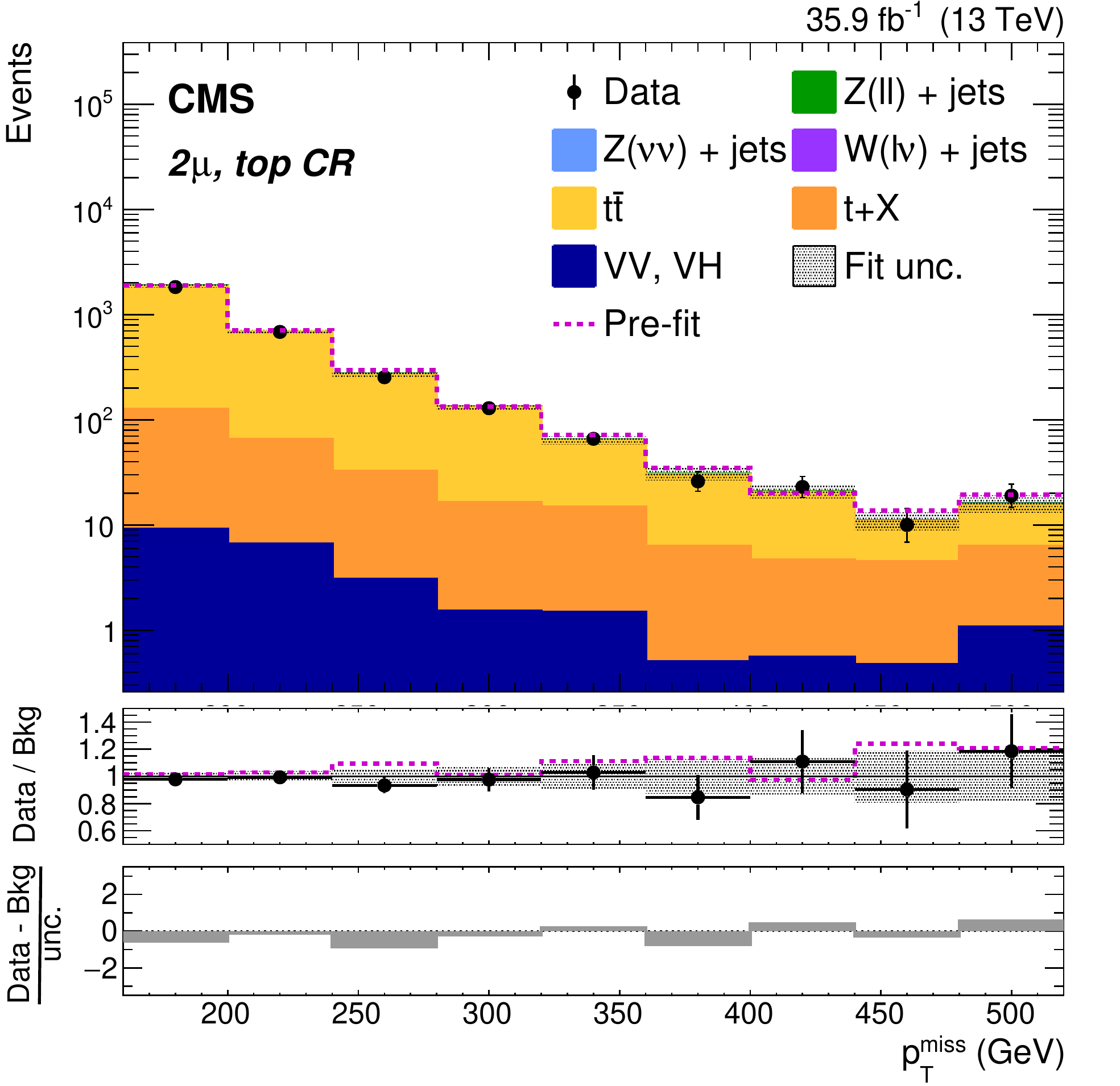}
\caption{Background-only post-fit \ptmiss distributions for the CRs of the SL selection. The total theory signal (\tDM and \ttDM summed together) is negligible and therefore is not shown. The last bin contains overflow events.
The dashed magenta lines show the total pre-fit background expectation in the upper panels, and the ratio of pre-fit total background to post-fit total background in the  middle panels. The lower panels show the difference between observed and post-fit total background divided by the full statistical and systematic uncertainties.
}
\label{fig:CR_SL_met_postfit}
\end{figure}

\begin{figure}[htbp!]
\centering
\includegraphics[width=0.42\textwidth]{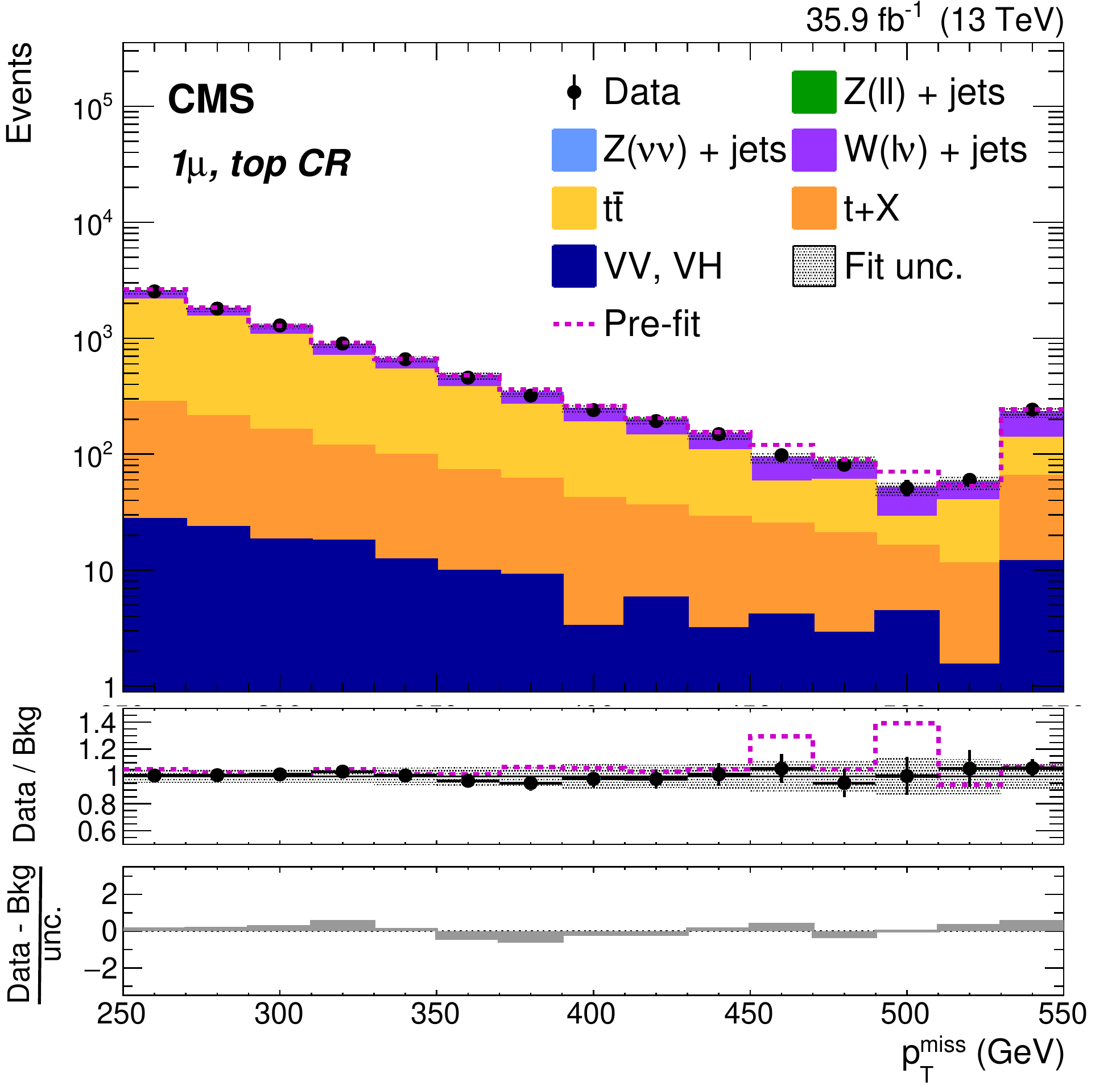}
\includegraphics[width=0.42\textwidth]{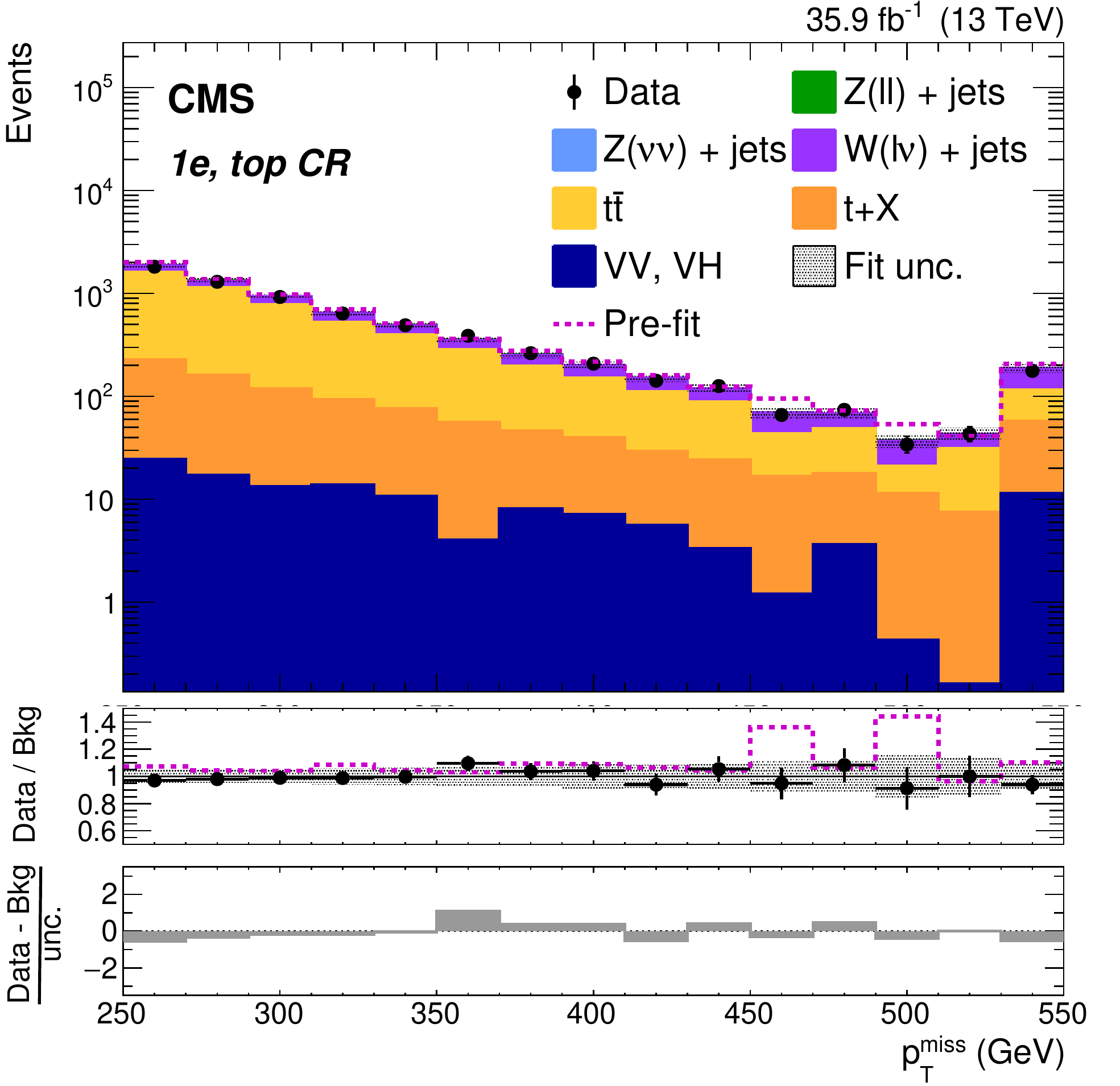}\\
\includegraphics[width=0.42\textwidth]{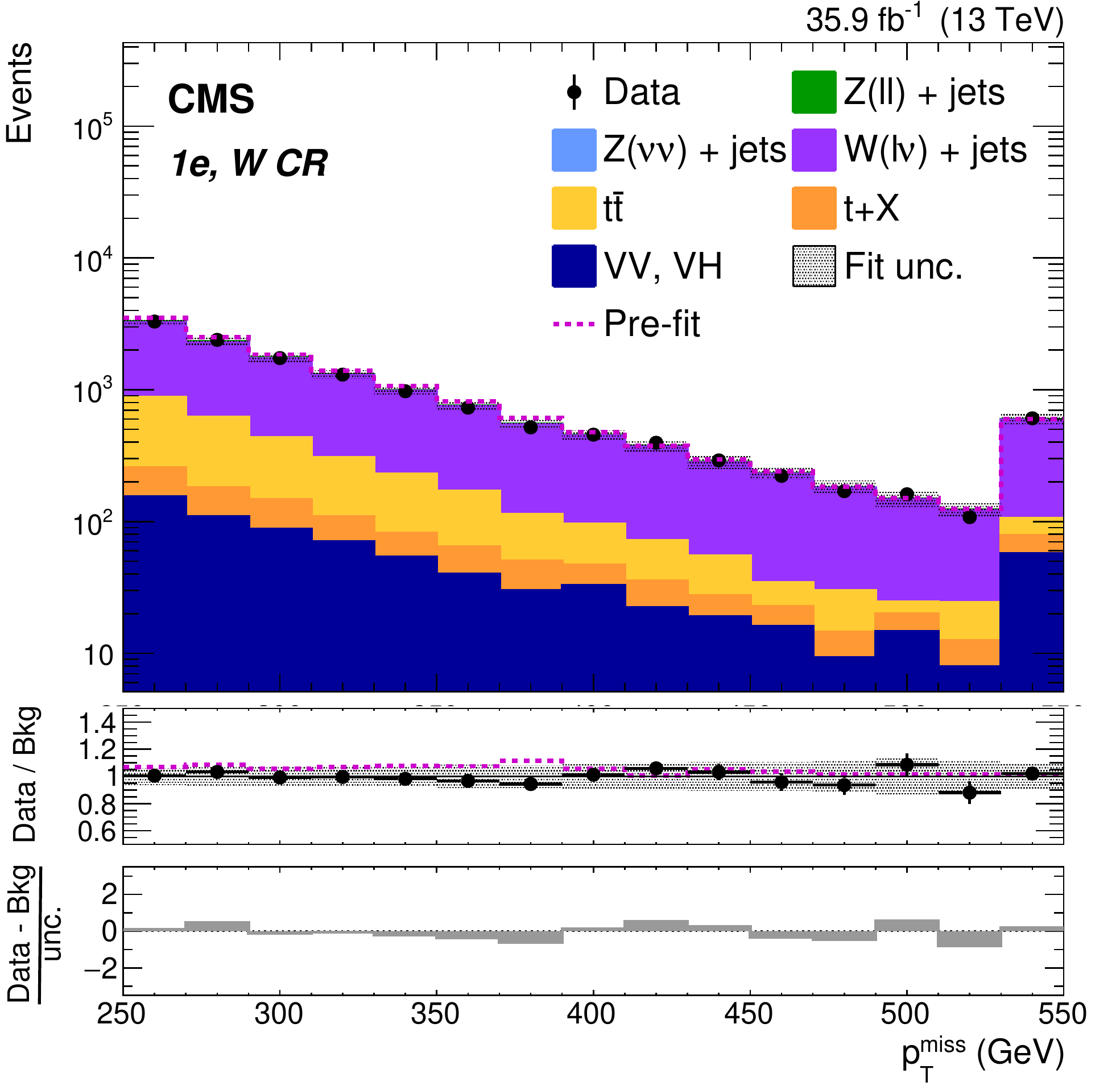}
\includegraphics[width=0.42\textwidth]{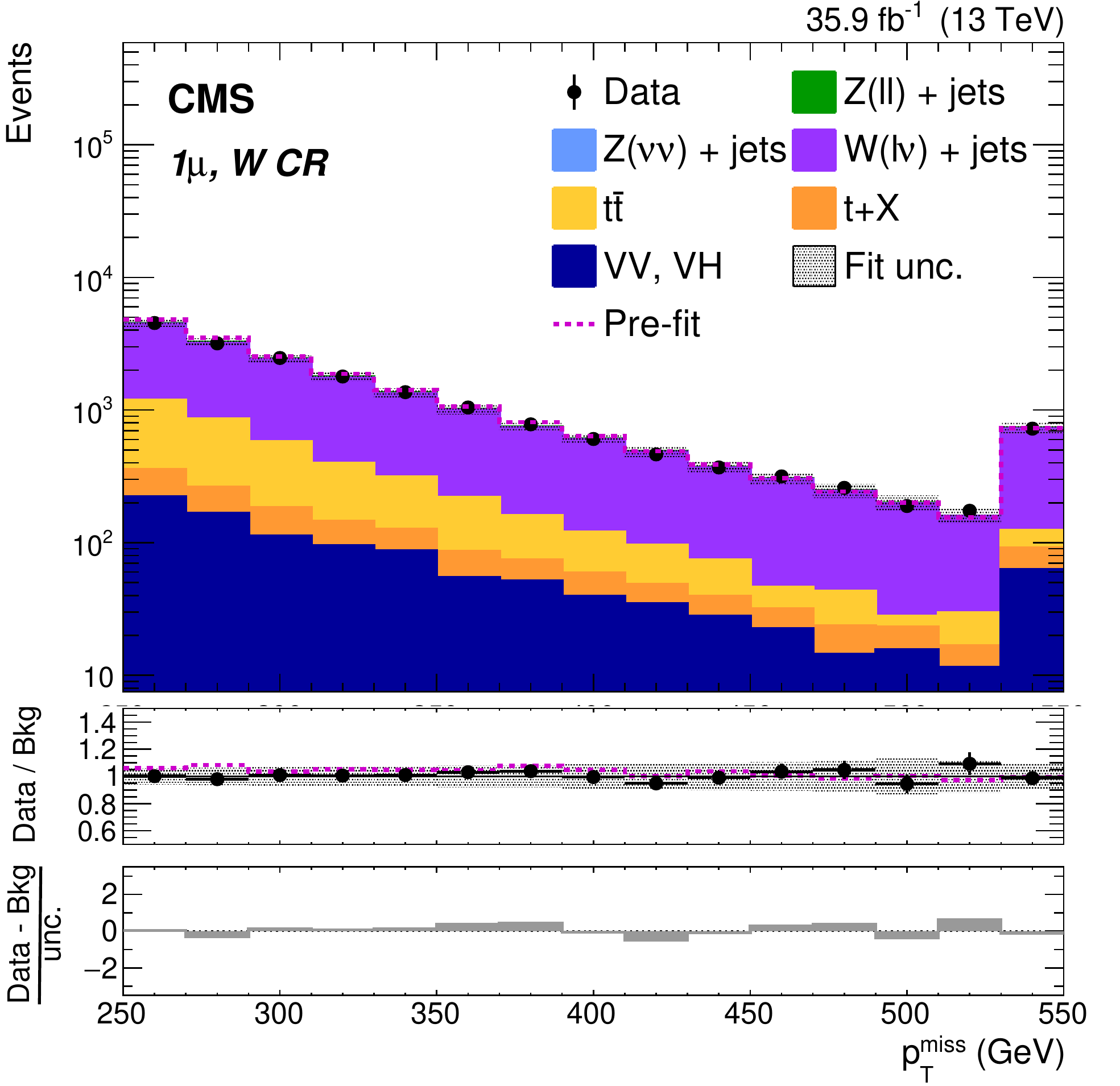}\\
\includegraphics[width=0.42\textwidth]{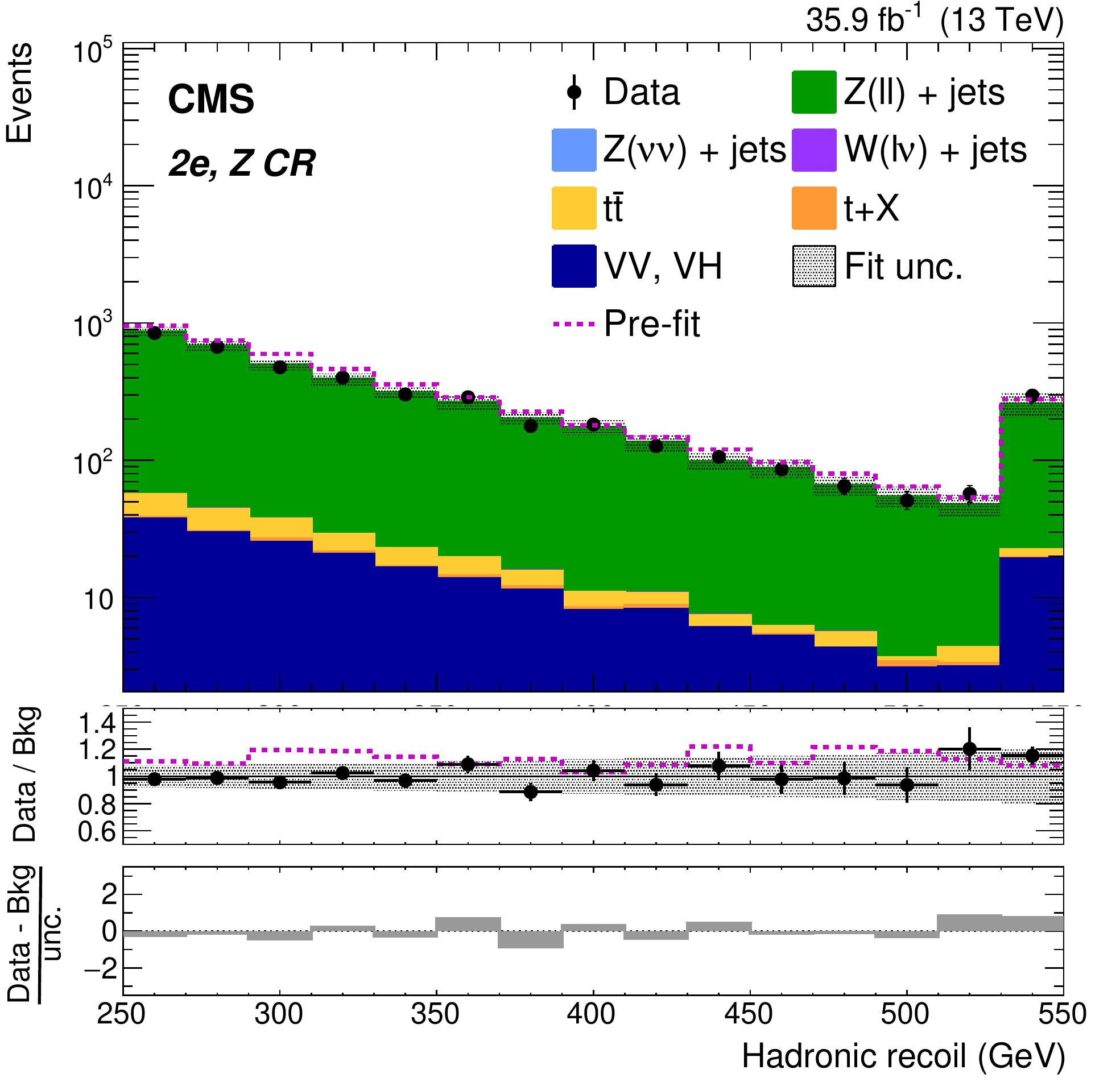}
\includegraphics[width=0.42\textwidth]{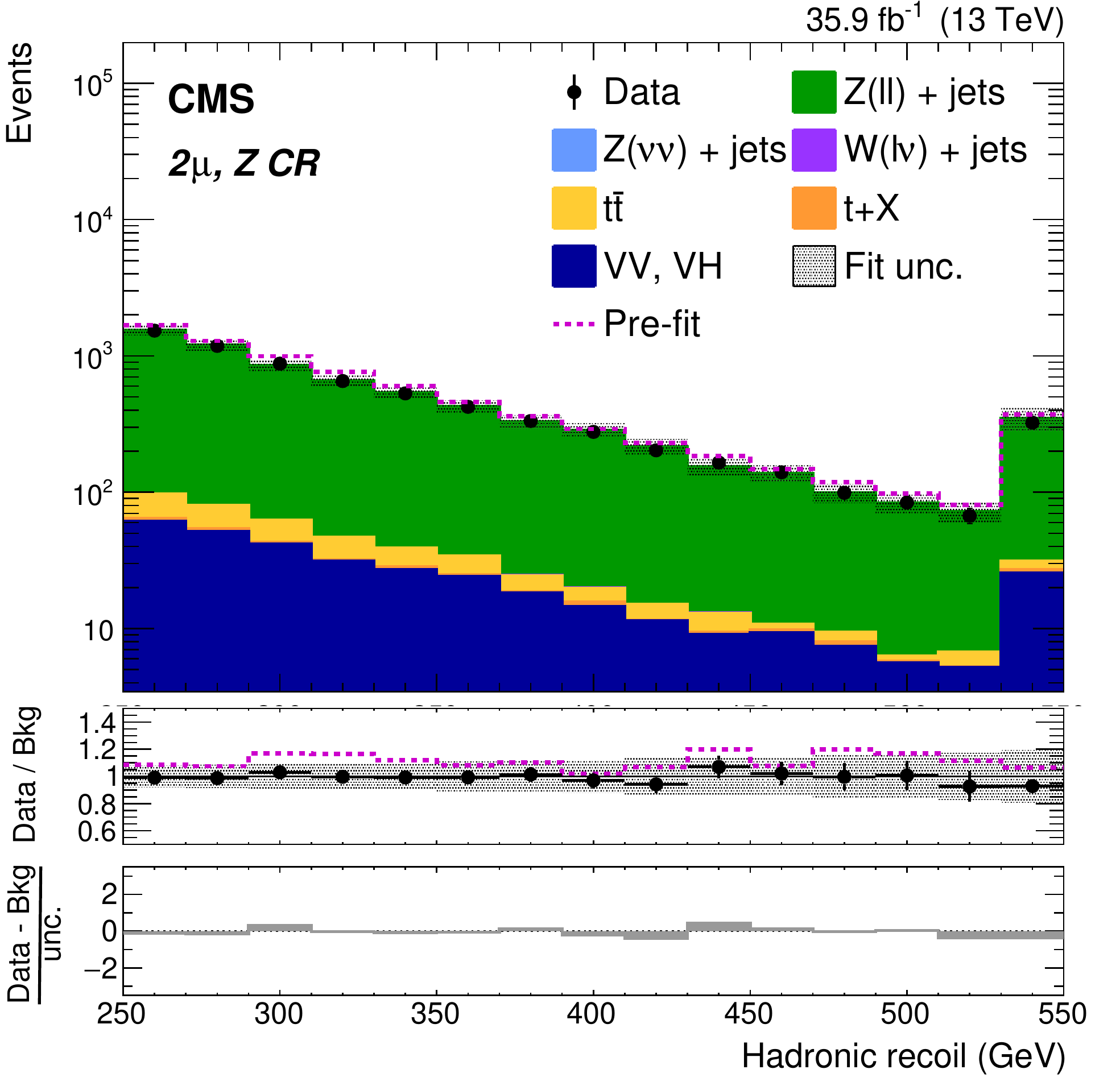}
\caption{Background-only post-fit \ptmiss distributions for the CRs of the AH selection. The total theory signal (\tDM and \ttDM summed together) is negligible and therefore is not shown. The last bin contains overflow events.
The dashed magenta lines show the total pre-fit background expectation in the upper panels, and the ratio of pre-fit total background to post-fit total background in the  middle panels. The lower panels show the difference between observed and post-fit total background divided by the full statistical and systematic uncertainties.
}
\label{fig:CR_HAD_met_postfit}
\end{figure}

\begin{figure}[htbp!]
\centering
\includegraphics[width=0.42\textwidth]{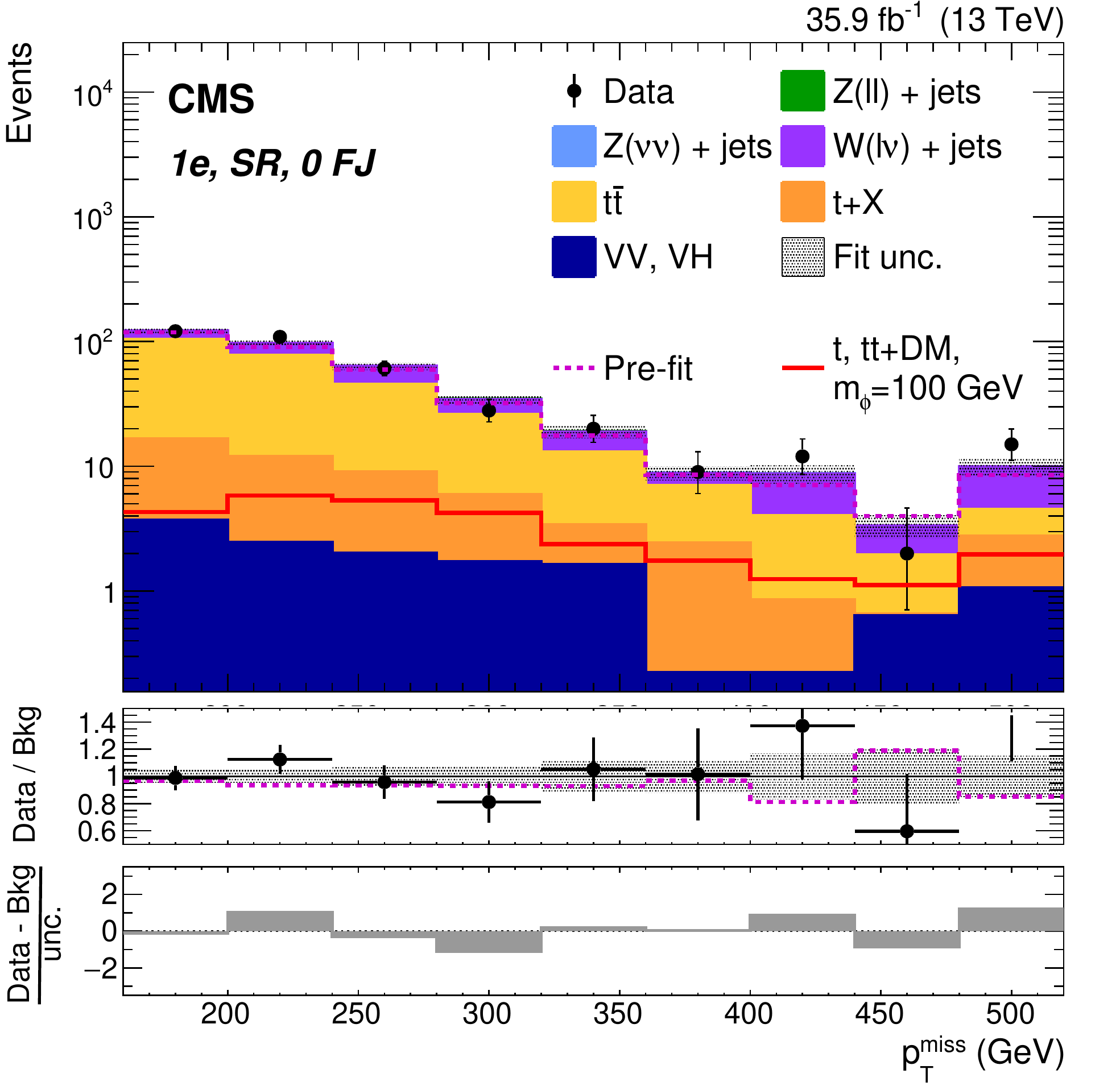}
\includegraphics[width=0.42\textwidth]{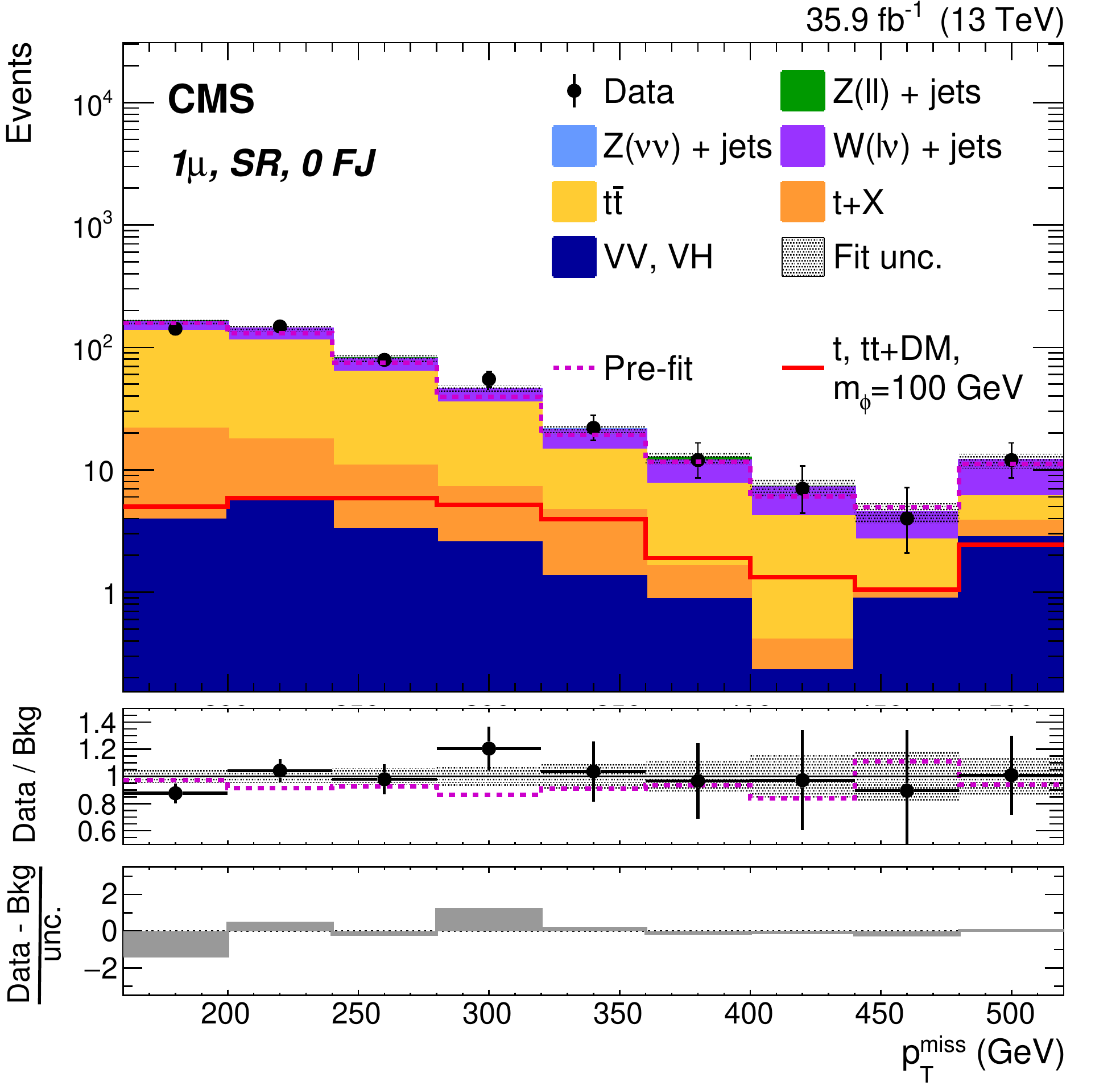}\\
\includegraphics[width=0.42\textwidth]{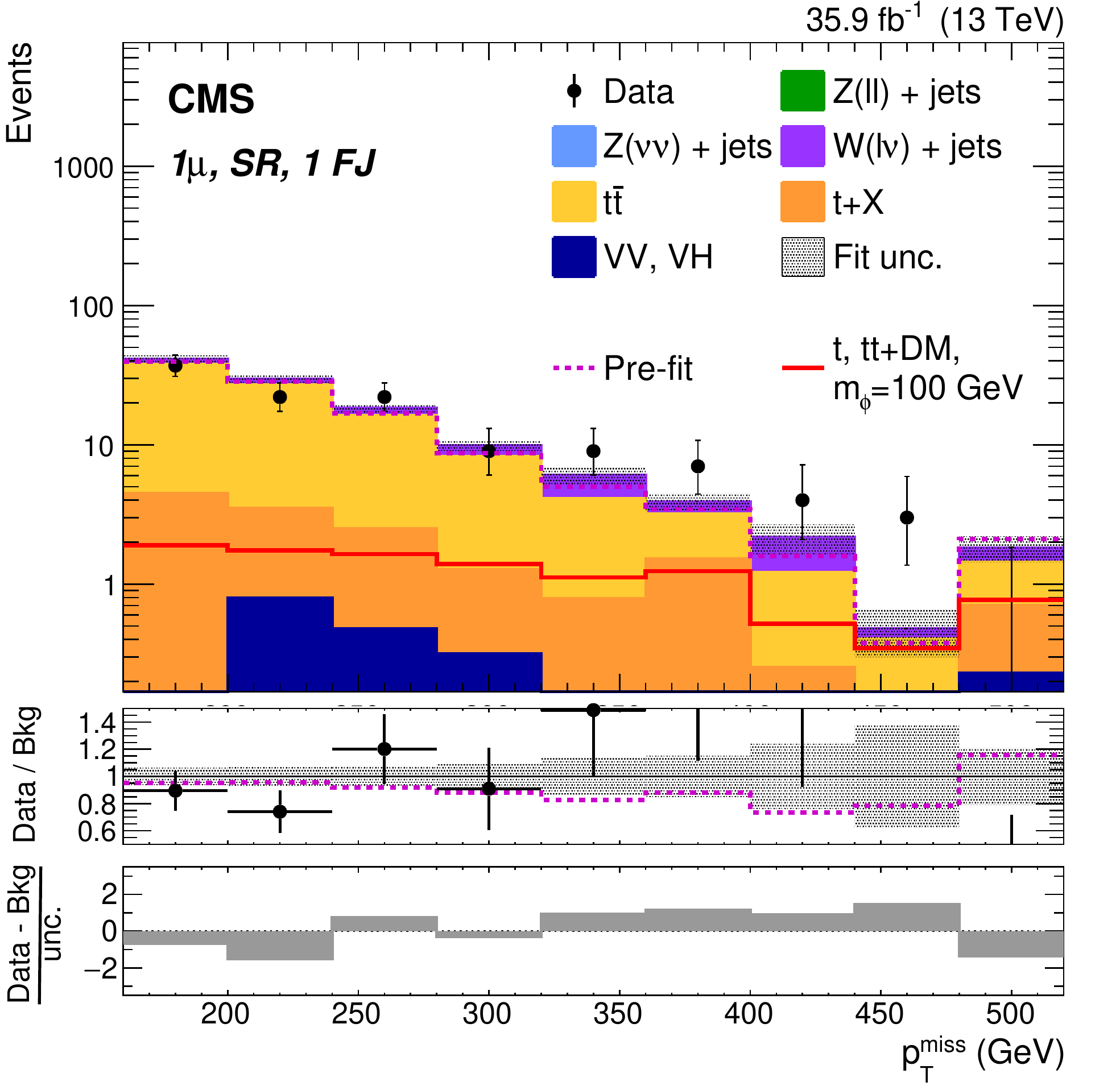}
\includegraphics[width=0.42\textwidth]{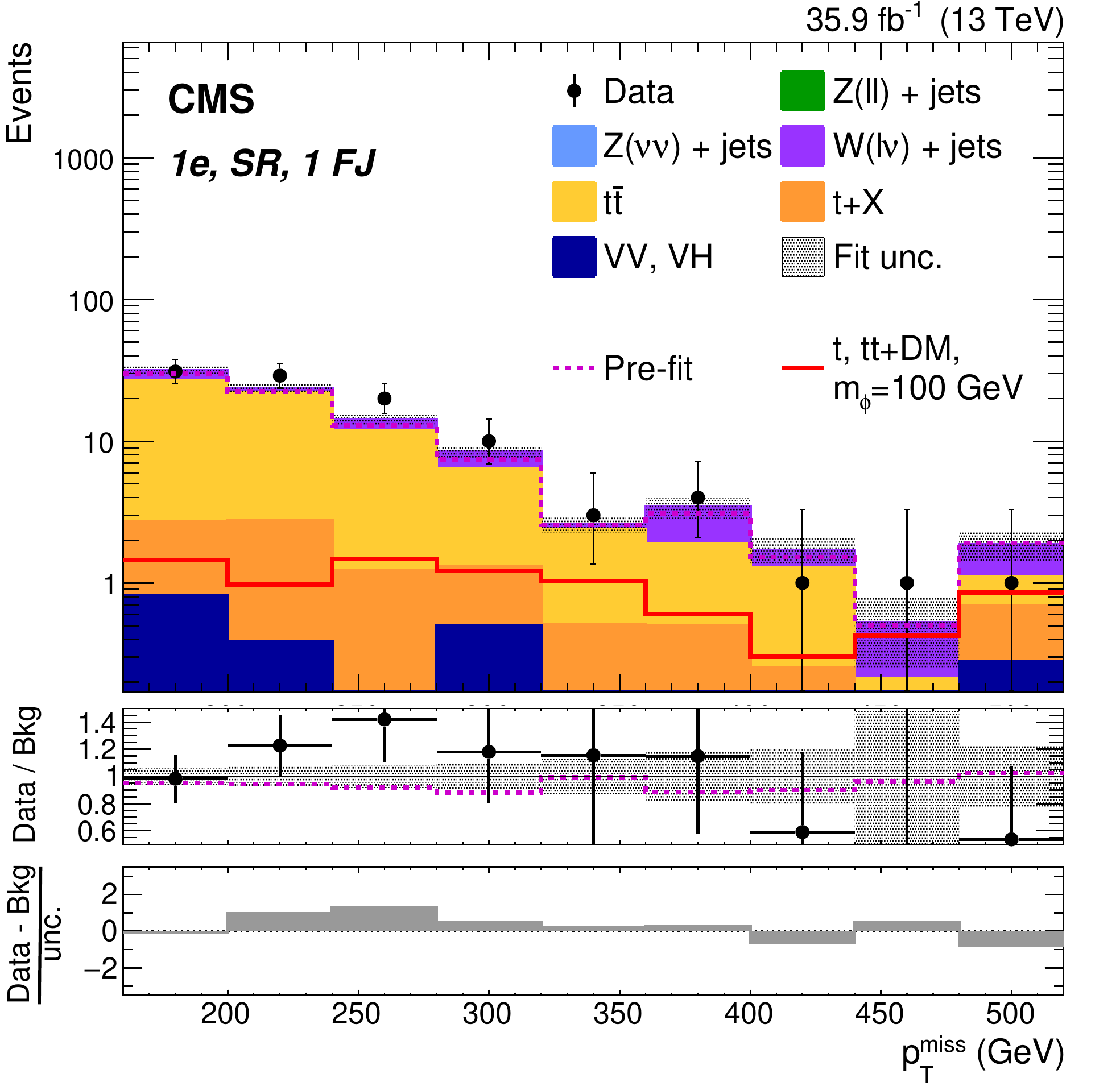}\\
\includegraphics[width=0.42\textwidth]{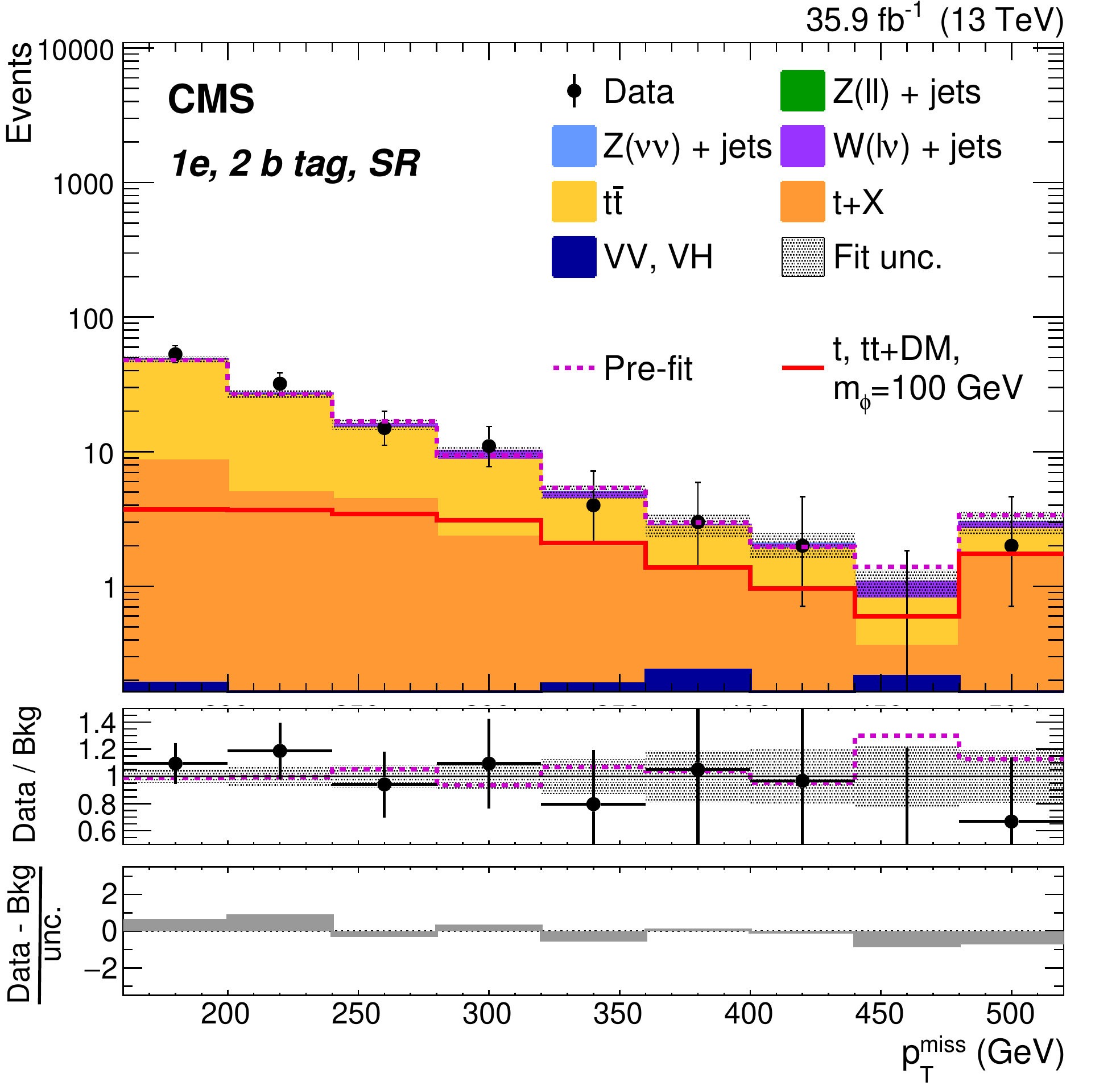}
\includegraphics[width=0.42\textwidth]{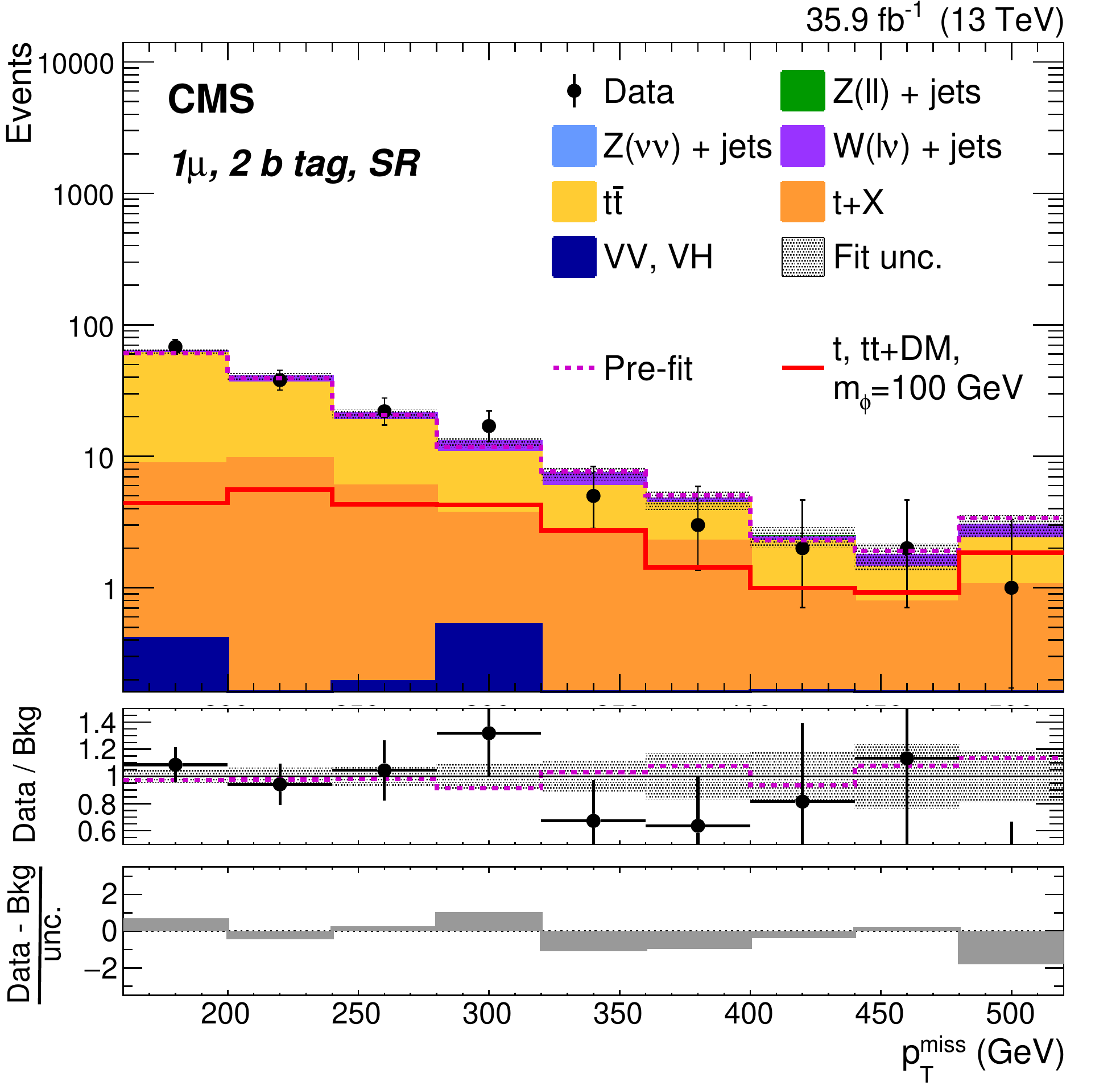}
\caption{Background-only post-fit \ptmiss distributions for the SRs of the SL selection. The total theory signal (\tDM and \ttDM summed together) is presented by the red solid lines for a scalar mediator mass of 100\GeV. The last bin contains overflow events.
The dashed magenta lines show the total pre-fit background expectation in the upper panels, and the ratio of pre-fit total background to post-fit total background in the  middle panels. The lower panels show the difference between observed and post-fit total background divided by the full statistical and systematic uncertainties.
}
\label{fig:SR_SL_met_postfit}
\end{figure}

\begin{figure}[htbp!]
\centering
\includegraphics[width=0.42\textwidth]{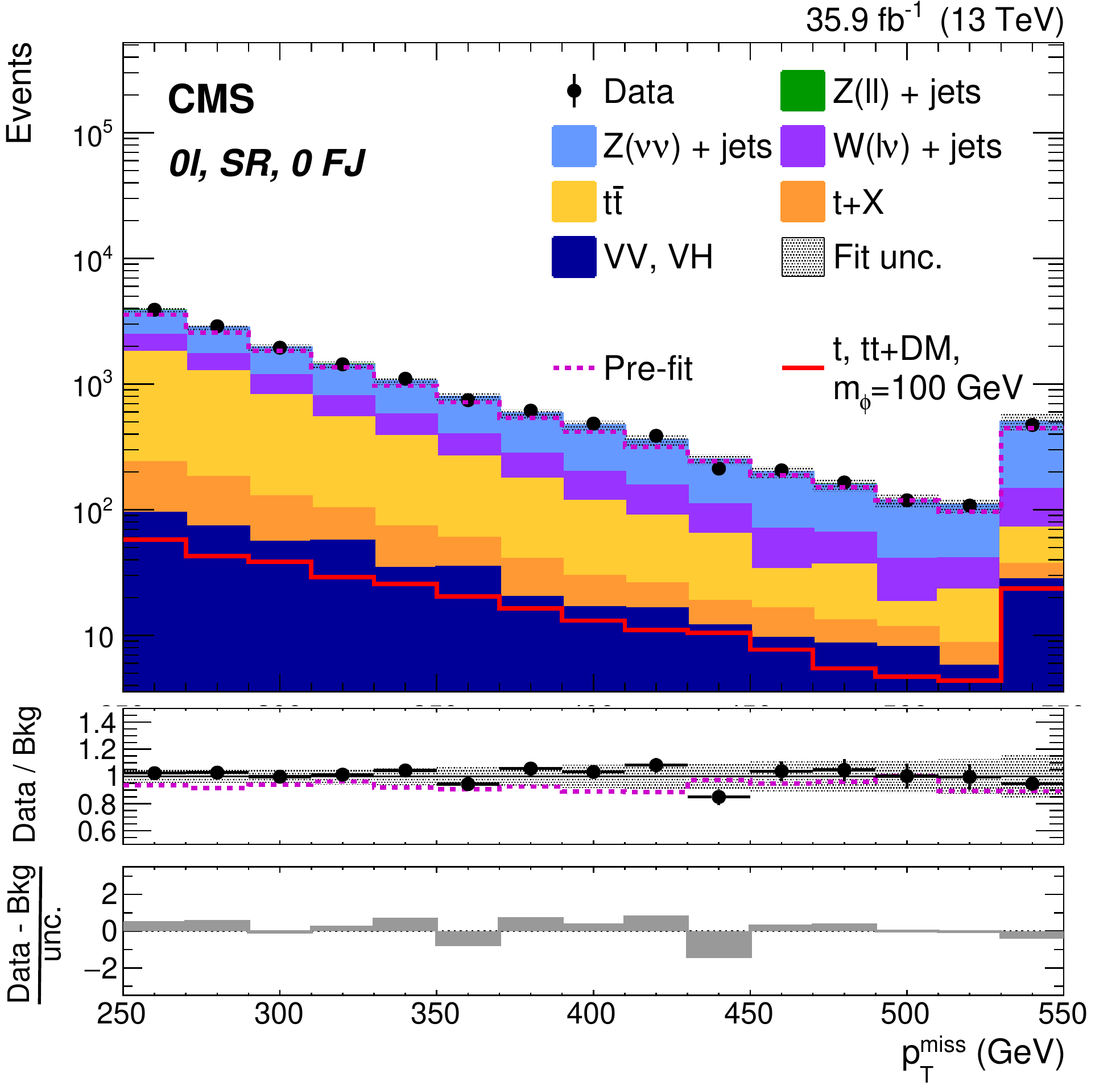}\\
\includegraphics[width=0.42\textwidth]{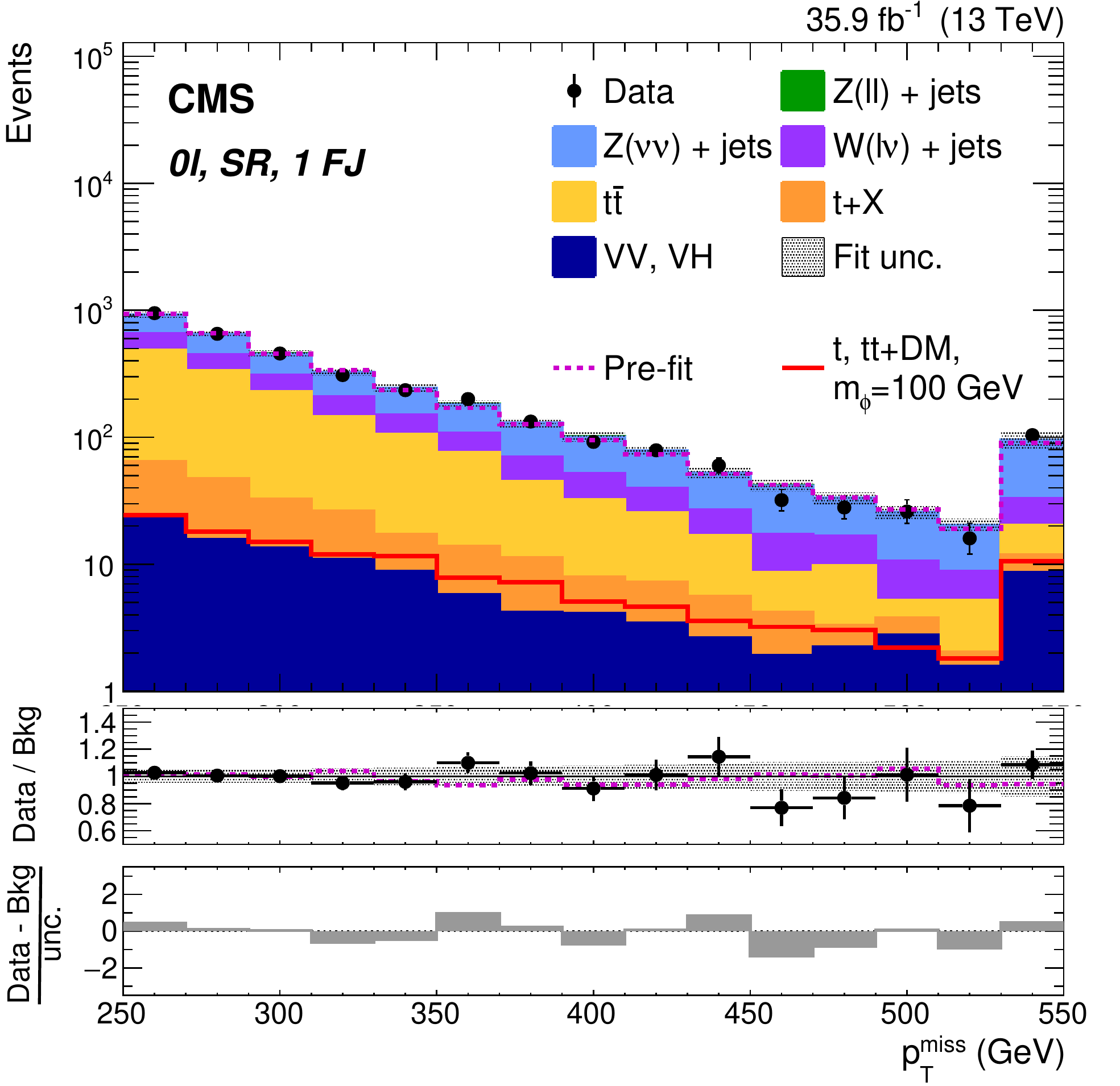}\\
\includegraphics[width=0.42\textwidth]{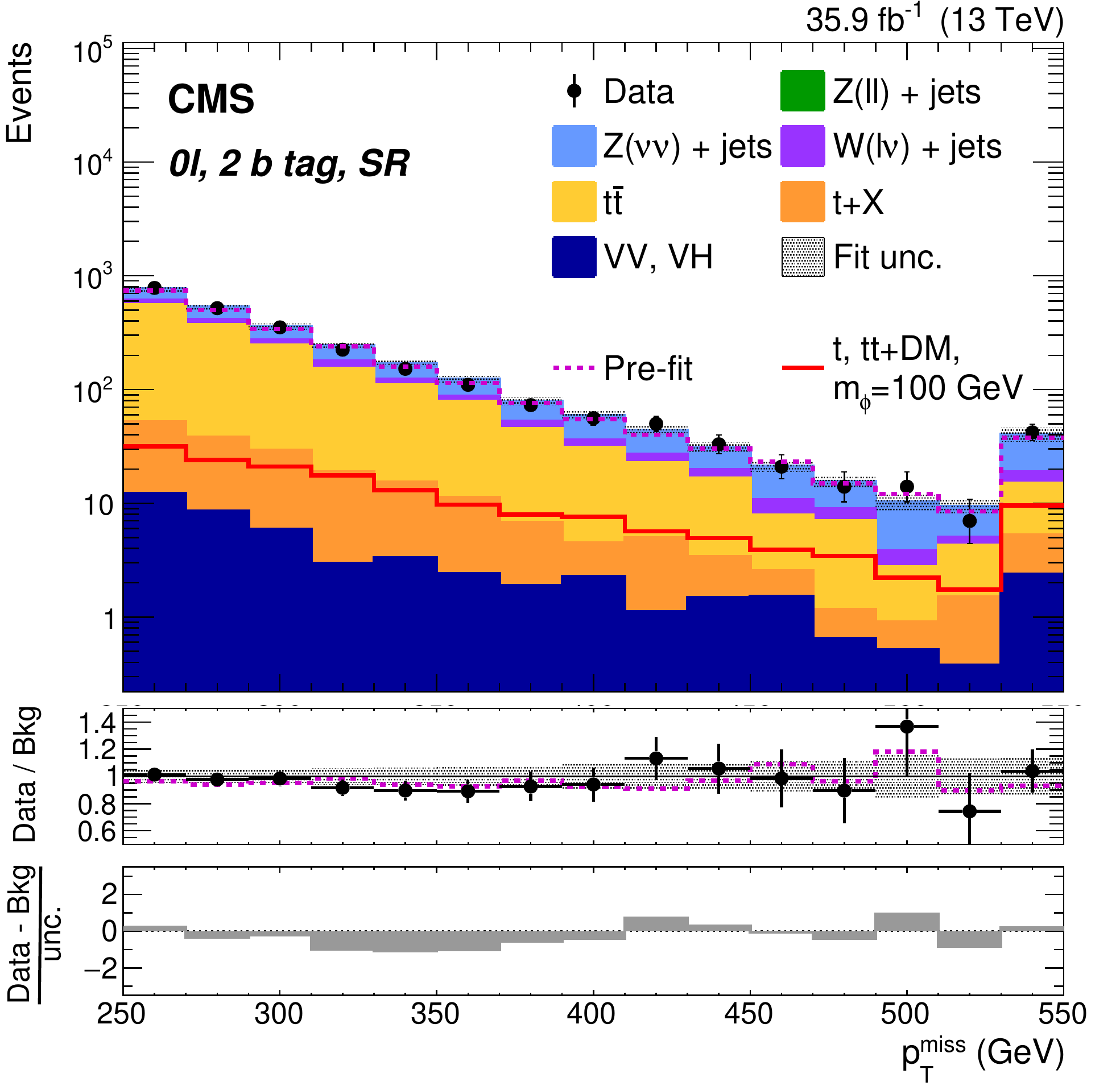}

\caption{Background-only post-fit \ptmiss distributions for the SRs of the AH selection. The total theory signal (\tDM and \ttDM summed together) is presented by the red solid lines for a scalar mediator mass of 100\GeV. The last bin contains overflow events.
The dashed magenta lines show the total pre-fit background expectation in the upper panels, and the ratio of pre-fit total background to post-fit total background in the  middle panels. The lower panels show the difference between observed and post-fit total background divided by the full statistical and systematic uncertainties.
}\label{fig:SR_HAD_met_postfit}
\end{figure}

The simultaneous fit to the binned \ptmiss distribution is performed combining the SL and AH regions.
The values for the background multiplicative factors extracted from the fit are on average close to one, with a root-mean-square deviation that ranges from 5\% to 21\%, depending on the background processes and on the category considered (SL or AH).
The post-fit distributions assuming the absence from the DM signal (\ie, the background only fit) are shown in Figs.~\ref{fig:CR_SL_met_postfit} and~\ref{fig:CR_HAD_met_postfit} (\ref{fig:SR_SL_met_postfit} and~\ref{fig:SR_HAD_met_postfit}) for the SL and AH CRs (SRs), respectively. No significant excess at high \ptmiss in the SRs is observed.
The SRs, both for the SL and AH channels, are divided into: 1 {\cPqb}-tagged jet and 0 forward jets, 1 {\cPqb}-tagged jet and ${\geq}1$ forward jets, and ${\geq}2$ {\cPqb}-tagged jets.
The plots also contain the pre-fit distributions, represented by the dashed magenta line.
The statistical and systematic uncertainties in the prediction are represented by hatched uncertainty bands, while the lower
 panels show the ratio of data and the post-fit prediction, and the bottom panels show the difference between the observed data events and the post-fit total background, divided by the full statistical and systematic uncertainty.

\section{Results}
\begin{figure}[htbp!]
\centering
\includegraphics[width=0.49\textwidth]{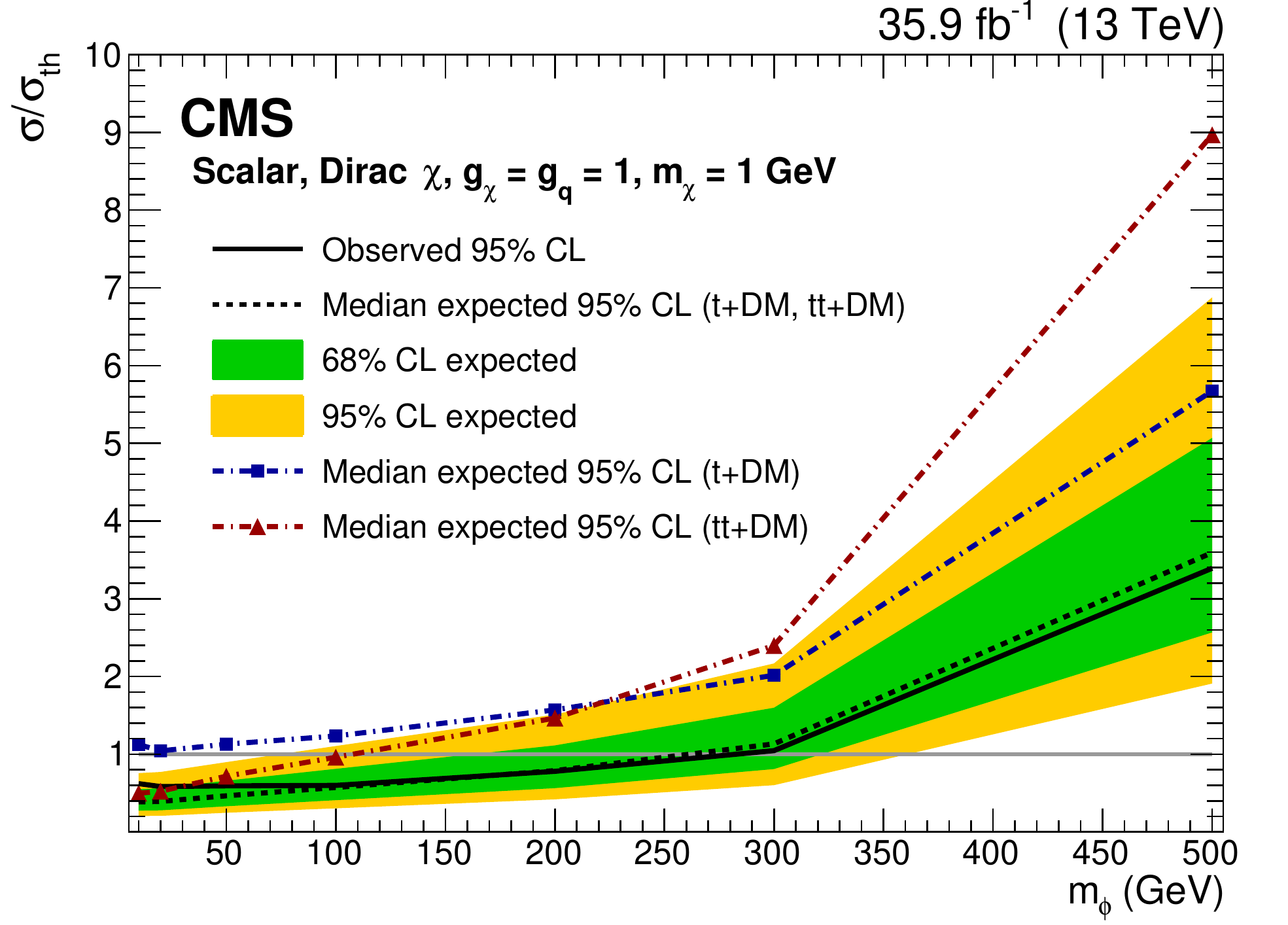}
\includegraphics[width=0.49\textwidth]{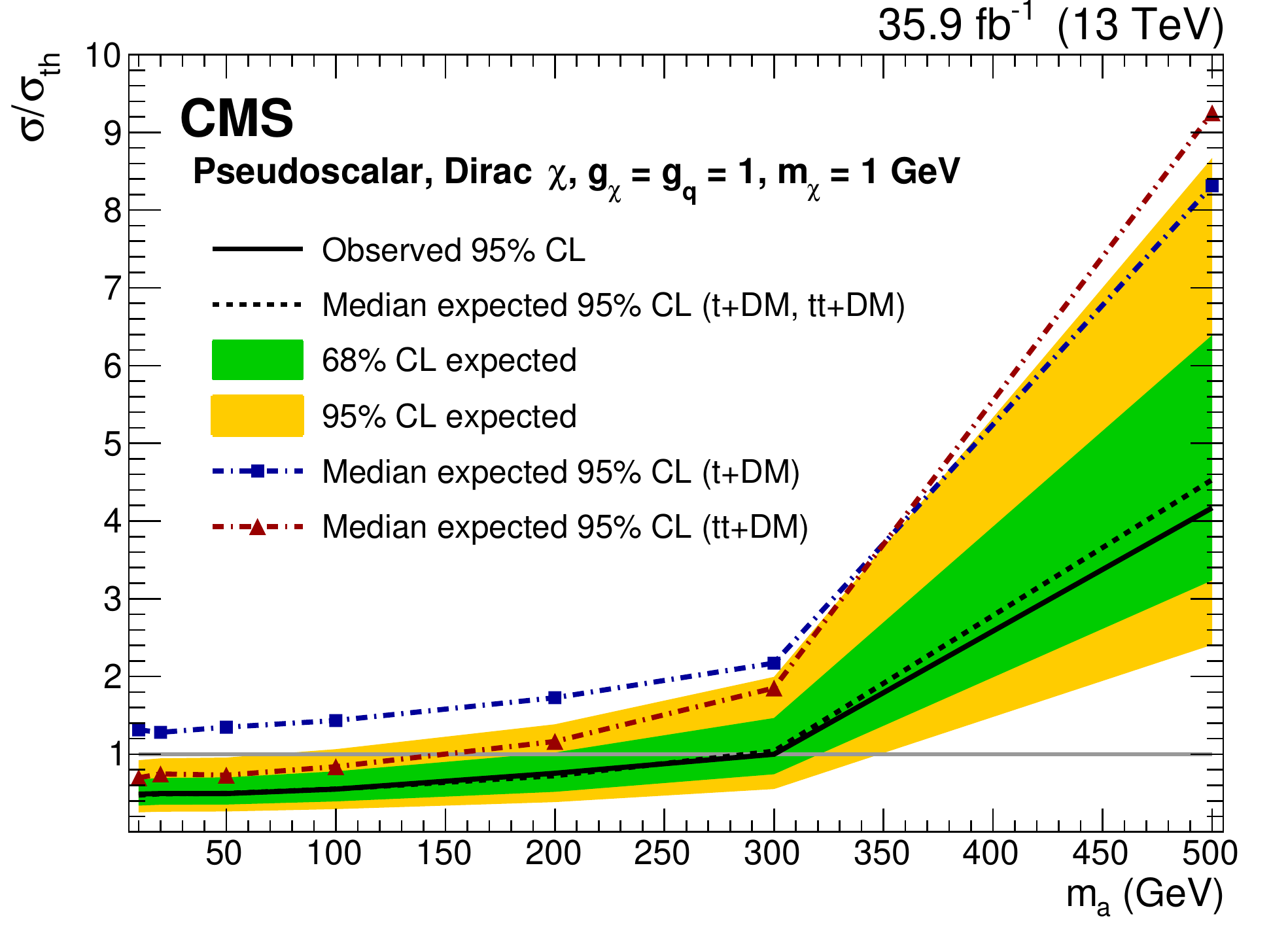}
\caption{
The expected and observed 95\% \CL limits on the DM production cross sections, relative to the theory predictions, shown for the scalar (left) and pseudoscalar (right) models. The expected limit for the \tDM signal alone is depicted by the blue dash-dotted line, while the \ttDM limit alone is given by the red dash-dotted line.
The observed limit on the sum of both signals is shown by the black solid line, while the expected value is shown by the black dashed line with the 68 and 95\% \CL uncertainty bands in green and yellow, respectively.
The solid horizontal line corresponds to $\sigma/\sigma_{\text{th}}=1$.}
\label{fig:limits_expected}
\end{figure}

Overall, data are found to be in agreement with the expected SM background in the SRs.
Upper limits at 95\% confidence level (\CL) are computed on the ratio between the measured and theoretical cross sections $\mu$, which is calculated with respect to the expected number of events for a scalar or pseudoscalar mediator and either the \tDM or \ttDM production modes separately, or summed together, where the results are referred to here as \tttDM. The theoretical cross sections for both signal models are obtained at LO.
The limits are calculated using a modified frequentist approach with a test statistic based on the profile likelihood in the asymptotic approximation and the \CLs criterion~\cite{JUNK1999435,cls,Cowan2011}.
We test different mediator mass scenarios with $\mchi=1\GeV$ and $g_{\cPq}=g_\chi=1$ and the results are shown in Fig.~\ref{fig:limits_expected} for scalar (left) and pseudoscalar (right) models.
 The expected limit for the \tDM signal alone is depicted by the blue dash-dotted line, while the expected \ttDM limit alone is given by
the red dash-dotted line.
The observed limit on the sum of both signals is represented by the black solid line, while its expected value is shown by the black dashed line with the 68 and 95\% \CL uncertainty bands in green and yellow, respectively.

For masses of the mediator particle below 200\GeV for the scalar model and below 300\GeV for the pseudoscalar model, the leading contribution
to the sensitivity of the analysis stems from \ttDM. This behavior is mostly driven by the larger cross section for the \ttDM process when compared to the sum of the
production processes for \tDM. However, the \tDM cross section drops less rapidly as a function of mediator particle mass in comparison to the \ttDM mode.
Additionally, the \ptmiss spectrum for a given mediator mass leans towards higher values for the \tDM signal model when compared to the \ttDM model.
These two features, together with the analysis specifically designed for both DM production modes and the statistical combination of the different SRs, lead up to a factor of two improvement at high mediator masses on the limits when compared to previous results~\cite{Sirunyan:2018dub}.
In particular, the ${\geq}1$ forward jet category, which is specifically designed to enhance \tDM $t$ channel events, improves the final results up to 14\%.

Table~\ref{tab:Limit} represents the final combined limits (SL + AH) for the
\tDM and \ttDM processes separately, and for the sum of the two processes.

Overall, we exclude mediator masses below 290 and 300\GeV for the scalar and pseudoscalar hypotheses, respectively.

\begin{table}[htbp]
\centering
\topcaption{Upper limits at 95\% \CL on the cross section ratio with respect to the expected DM signal for different scalar ($\phi$) or pseudoscalar (\Pa) mediator masses, $\mchi=1\GeV$, and $g_{\chi}=g_{\cPq}=1$ for the combination of SL and AH signal regions. The median expected value and its 68 and 95\% confidence intervals (CIs) are given.}
\begin{tabular}{ccccccccccccc}
\hline
& \multicolumn{1}{c}{} & & \multicolumn{2}{c}{\tDM} & & \multicolumn{2}{c}{\ttDM} &  &\multicolumn{4}{l}{\tttDM sum}  \\ \cline{4-5}\cline{7-8}\cline{10-13}
& \multicolumn{1}{c}{ ${m}_{\phi/\Pa}$(\GeVns{})} & & Obs. & Exp. & & Obs. & Exp. & & Obs. & Exp. &  68\% CI & 95\% CI \\ \hline
\multirow{7}{*}{\rotatebox[origin=c]{90}{Scalar}}
& 10 & & 1.59 & 1.12 & & 0.91 & 0.50 & & 0.62 & 0.39 & $\left[0.27,\:0.55\right]$ & $\left[0.21,\:0.76\right]$\\
& 20 & & 1.38 & 1.04 & & 0.84 & 0.52 & & 0.58 & 0.39 & $\left[0.28,\:0.56\right]$ & $\left[0.21,\:0.77\right]$\\
& 50 & & 1.15 & 1.13 & & 1.11 & 0.72 & & 0.59 & 0.46 & $\left[0.33,\:0.66\right]$ & $\left[0.25,\:0.90\right]$\\
& 100 & & 1.43 & 1.23 & & 0.94 & 0.96 & & 0.60 & 0.57 & $\left[0.41,\:0.81\right]$ & $\left[0.30,\:1.11\right]$\\
& 200 & & 1.66 & 1.57 & & 1.37 & 1.46 & & 0.78 & 0.79 & $\left[0.56,\:1.11\right]$ & $\left[0.42,\:1.51\right]$\\
& 300 & & 1.97 & 2.02 & & 2.09 & 2.40 & & 1.05 & 1.13 & $\left[0.81,\:1.60\right]$ & $\left[0.60,\:2.17\right]$\\
& 500 & & 5.84 & 5.67 & & 7.48 & 8.97 & & 3.39 & 3.59 & $\left[2.57,\:5.07\right]$ & $\left[1.91,\:6.88\right]$\\
[\cmsTabSkip]
\multirow{7}{*}{\rotatebox[origin=c]{90}{Pseudoscalar}}
& 10 & & 1.43 & 1.31 & & 0.70 & 0.70 & & 0.49 & 0.47 & $\left[0.34,\:0.67\right]$ & $\left[0.25,\:0.92\right]$\\
& 20 & & 1.43 & 1.28 & & 0.71 & 0.75 & & 0.49 & 0.49 & $\left[0.35,\:0.70\right]$ & $\left[0.26,\:0.95\right]$\\
& 50 & & 1.48 & 1.35 & & 0.70 & 0.73 & & 0.49 & 0.50 & $\left[0.35,\:0.70\right]$ & $\left[0.26,\:0.96\right]$\\
& 100 & & 1.53 & 1.43 & & 0.81 & 0.84 & & 0.55 & 0.55 & $\left[0.39,\:0.78\right]$ & $\left[0.29,\:1.06\right]$\\
& 200 & & 1.89 & 1.73 & & 1.18 & 1.16 & & 0.76 & 0.72 & $\left[0.52,\:1.02\right]$ & $\left[0.38,\:1.38\right]$\\
& 300 & & 2.17 & 2.17 & & 1.74 & 1.85 & & 1.00 & 1.04 & $\left[0.74,\:1.47\right]$ & $\left[0.55,\:2.00\right]$\\
& 500 & & 8.22 & 8.31 & & 8.00 & 9.25 & & 4.17 & 4.53 & $\left[3.24,\:6.39\right]$ & $\left[2.41,\:8.67\right]$\\
\hline
\end{tabular}
\label{tab:Limit}
\end{table}

\section {Summary}

The first search at the LHC for dark matter (DM) produced in association with a single top quark or a top quark pair in interactions mediated by a neutral scalar or pseudoscalar particle in proton-proton collisions at a center-of-mass energy of 13\TeV has been presented. The data correspond to an integrated luminosity of 35.9\fbinv recorded by the CMS experiment in 2016.
No significant deviations with respect to standard model predictions are observed and the results are interpreted in the context of a simplified model in which a scalar or pseudoscalar mediator particle couples to the top quark and subsequently decays into two DM particles.

Scalar and pseudoscalar mediator masses below 290 and 300\GeV are excluded at 95\% confidence level assuming a DM particle mass of 1\GeV and mediator couplings to fermions and DM particles equal to unity.
This analysis provides the most stringent limits derived at the LHC for these new spin-0 mediator particles.

\begin{acknowledgments}
We congratulate our colleagues in the CERN accelerator departments for the excellent performance of the LHC and thank the technical and administrative staffs at CERN and at other CMS institutes for their contributions to the success of the CMS effort. In addition, we gratefully acknowledge the computing centres and personnel of the Worldwide LHC Computing Grid for delivering so effectively the computing infrastructure essential to our analyses. Finally, we acknowledge the enduring support for the construction and operation of the LHC and the CMS detector provided by the following funding agencies: BMBWF and FWF (Austria); FNRS and FWO (Belgium); CNPq, CAPES, FAPERJ, FAPERGS, and FAPESP (Brazil); MES (Bulgaria); CERN; CAS, MoST, and NSFC (China); COLCIENCIAS (Colombia); MSES and CSF (Croatia); RPF (Cyprus); SENESCYT (Ecuador); MoER, ERC IUT, and ERDF (Estonia); Academy of Finland, MEC, and HIP (Finland); CEA and CNRS/IN2P3 (France); BMBF, DFG, and HGF (Germany); GSRT (Greece); NKFIA (Hungary); DAE and DST (India); IPM (Iran); SFI (Ireland); INFN (Italy); MSIP and NRF (Republic of Korea); MES (Latvia); LAS (Lithuania); MOE and UM (Malaysia); BUAP, CINVESTAV, CONACYT, LNS, SEP, and UASLP-FAI (Mexico); MOS (Montenegro); MBIE (New Zealand); PAEC (Pakistan); MSHE and NSC (Poland); FCT (Portugal); JINR (Dubna); MON, RosAtom, RAS, RFBR, and NRC KI (Russia); MESTD (Serbia); SEIDI, CPAN, PCTI, and FEDER (Spain); MOSTR (Sri Lanka); Swiss Funding Agencies (Switzerland); MST (Taipei); ThEPCenter, IPST, STAR, and NSTDA (Thailand); TUBITAK and TAEK (Turkey); NASU and SFFR (Ukraine); STFC (United Kingdom); DOE and NSF (USA).

\hyphenation{Rachada-pisek} Individuals have received support from the Marie-Curie programme and the European Research Council and Horizon 2020 Grant, contract No. 675440 (European Union); the Leventis Foundation; the A.P.\ Sloan Foundation; the Alexander von Humboldt Foundation; the Belgian Federal Science Policy Office; the Fonds pour la Formation \`a la Recherche dans l'Industrie et dans l'Agriculture (FRIA-Belgium); the Agentschap voor Innovatie door Wetenschap en Technologie (IWT-Belgium); the F.R.S.-FNRS and FWO (Belgium) under the ``Excellence of Science -- EOS'' -- be.h project n.\ 30820817; the Ministry of Education, Youth and Sports (MEYS) of the Czech Republic; the Lend\"ulet (``Momentum'') Programme and the J\'anos Bolyai Research Scholarship of the Hungarian Academy of Sciences, the New National Excellence Program \'UNKP, the NKFIA research grants 123842, 123959, 124845, 124850, and 125105 (Hungary); the Council of Science and Industrial Research, India; the HOMING PLUS programme of the Foundation for Polish Science, cofinanced from European Union, Regional Development Fund, the Mobility Plus programme of the Ministry of Science and Higher Education, the National Science Center (Poland), contracts Harmonia 2014/14/M/ST2/00428, Opus 2014/13/B/ST2/02543, 2014/15/B/ST2/03998, and 2015/19/B/ST2/02861, Sonata-bis 2012/07/E/ST2/01406; the National Priorities Research Program by Qatar National Research Fund; the Programa Estatal de Fomento de la Investigaci{\'o}n Cient{\'i}fica y T{\'e}cnica de Excelencia Mar\'{\i}a de Maeztu, grant MDM-2015-0509 and the Programa Severo Ochoa del Principado de Asturias; the Thalis and Aristeia programmes cofinanced by EU-ESF and the Greek NSRF; the Rachadapisek Sompot Fund for Postdoctoral Fellowship, Chulalongkorn University and the Chulalongkorn Academic into Its 2nd Century Project Advancement Project (Thailand); the Welch Foundation, contract C-1845; and the Weston Havens Foundation (USA).
\end{acknowledgments}

\bibliography{auto_generated}

\cleardoublepage \appendix\section{The CMS Collaboration \label{app:collab}}\begin{sloppypar}\hyphenpenalty=5000\widowpenalty=500\clubpenalty=5000\vskip\cmsinstskip
\textbf{Yerevan Physics Institute, Yerevan, Armenia}\\*[0pt]
A.M.~Sirunyan, A.~Tumasyan
\vskip\cmsinstskip
\textbf{Institut f\"{u}r Hochenergiephysik, Wien, Austria}\\*[0pt]
W.~Adam, F.~Ambrogi, E.~Asilar, T.~Bergauer, J.~Brandstetter, M.~Dragicevic, J.~Er\"{o}, A.~Escalante~Del~Valle, M.~Flechl, R.~Fr\"{u}hwirth\cmsAuthorMark{1}, V.M.~Ghete, J.~Hrubec, M.~Jeitler\cmsAuthorMark{1}, N.~Krammer, I.~Kr\"{a}tschmer, D.~Liko, T.~Madlener, I.~Mikulec, N.~Rad, H.~Rohringer, J.~Schieck\cmsAuthorMark{1}, R.~Sch\"{o}fbeck, M.~Spanring, D.~Spitzbart, W.~Waltenberger, J.~Wittmann, C.-E.~Wulz\cmsAuthorMark{1}, M.~Zarucki
% ...
\vskip\cmsinstskip
% ...
1:  Also at Vienna University of Technology, Vienna, Austria\\
% ...
\end{sloppypar}
\end{document}